\input harvmac
\input amssym
\input epsf


\newfam\frakfam
\font\teneufm=eufm10
\font\seveneufm=eufm7
\font\fiveeufm=eufm5
\textfont\frakfam=\teneufm
\scriptfont\frakfam=\seveneufm
\scriptscriptfont\frakfam=\fiveeufm




\newfam\dsromfam
\font\tendsrom=dsrom10
\textfont\dsromfam=\tendsrom
\def\ds{\fam\dsromfam \tendsrom}


\newfam\mbffam
\font\tenmbf=cmmib10
\font\sevenmbf=cmmib7
\font\fivembf=cmmib5
\textfont\mbffam=\tenmbf
\scriptfont\mbffam=\sevenmbf
\scriptscriptfont\mbffam=\fivembf


\newfam\mbfcalfam
\font\tenmbfcal=cmbsy10
\font\sevenmbfcal=cmbsy7
\font\fivembfcal=cmbsy5
\textfont\mbfcalfam=\tenmbfcal
\scriptfont\mbfcalfam=\sevenmbfcal
\scriptscriptfont\mbfcalfam=\fivembfcal


\newfam\mscrfam
\font\tenmscr=rsfs10
\font\sevenmscr=rsfs7
\font\fivemscr=rsfs5
\textfont\mscrfam=\tenmscr
\scriptfont\mscrfam=\sevenmscr
\scriptscriptfont\mscrfam=\fivemscr





\def\bar{\overline}
\def\b{\bar}
\def\bsq#1{{{\b{#1}}^{\lower 2.5pt\hbox{$\scriptstyle 2$}}}}
\def\bexp#1#2{{{\b{#1}}^{\lower 2.5pt\hbox{$\scriptstyle #2$}}}}
\def\dotexp#1#2{{{#1}^{\lower 2.5pt\hbox{$\scriptstyle #2$}}}}


\def\rt2{\sqrt{2}}

\def\grad{\nabla}

\def\Tr{\mathop{\rm Tr}}

\def\lrpar#1{\raise1.5ex\hbox{$\leftrightarrow$}\mkern-16.5mu \partial #1}


\font\tenbifull=cmmib10
\font\tenbimed=cmmib7
\font\tenbismall=cmmib5
\textfont9=\tenbifull \scriptfont9=\tenbimed
\scriptscriptfont9=\tenbismall

\mathchardef\bbGamma="7000
\mathchardef\bbDelta="7001
\mathchardef\bbPhi="7002
\mathchardef\bbAlpha="7003
\mathchardef\bbXi="7004
\mathchardef\bbPi="7005
\mathchardef\bbSigma="7006
\mathchardef\bbUpsilon="7007
\mathchardef\bbTheta="7008
\mathchardef\bbPsi="7009
\mathchardef\bbOmega="700A
\mathchardef\bbalpha="710B
\mathchardef\bbbeta="710C
\mathchardef\bbgamma="710D
\mathchardef\bbdelta="710E
\mathchardef\bbepsilon="710F
\mathchardef\bbzeta="7110
\mathchardef\bbeta="7111
\mathchardef\bbtheta="7112
\mathchardef\bbiota="7113
\mathchardef\bbkappa="7114
\mathchardef\bblambda="7115
\mathchardef\bbmu="7116
\mathchardef\bbnu="7117
\mathchardef\bbxi="7118
\mathchardef\bbpi="7119
\mathchardef\bbrho="711A
\mathchardef\bbsigma="711B
\mathchardef\bbtau="711C
\mathchardef\bbupsilon="711D
\mathchardef\bbphi="711E
\mathchardef\bbchi="711F
\mathchardef\bbpsi="7120
\mathchardef\bbomega="7121
\mathchardef\bbvarepsilon="7122
\mathchardef\bbvartheta="7123
\mathchardef\bbvarpi="7124
\mathchardef\bbvarrho="7125
\mathchardef\bbvarsigma="7126
\mathchardef\bbvarphi="7127





\def\CC{{\cal C}}

\def\CF{{\cal F}}

\def\CH{{\cal H}}

\def\CK{{\cal K}}
\def\CL{{\cal L}}
\def\CM{{\cal M}}
\def\CN{{\cal N}}
\def\CO{{\cal O}}

\def\CR{{\cal R}}


\def\1{{\ds 1}}


\def\>{\rangle}
\def\<{\langle}

\def\slashchar#1{\setbox0=\hbox{$#1$}           
   \dimen0=\wd0                                 
   \setbox1=\hbox{/} \dimen1=\wd1               
   \ifdim\dimen0>\dimen1                        
      \rlap{\hbox to \dimen0{\hfil/\hfil}}      
      #1                                        
   \else                                        
      \rlap{\hbox to \dimen1{\hfil$#1$\hfil}}   
      /                                         
   \fi}

\lref\RattazziPE{
  R.~Rattazzi, V.~S.~Rychkov, E.~Tonni and A.~Vichi,
  ``Bounding scalar operator dimensions in 4D CFT,''
  JHEP {\bf 0812}, 031 (2008)
  [arXiv:0807.0004 [hep-th]].
}

\lref\RychkovIJ{
  V.~S.~Rychkov and A.~Vichi,
  ``Universal Constraints on Conformal Operator Dimensions,''
  Phys.\ Rev.\  D {\bf 80}, 045006 (2009)
  [arXiv:0905.2211 [hep-th]].
}

\lref\RattazziYC{
  R.~Rattazzi, S.~Rychkov and A.~Vichi,
  ``Bounds in 4D Conformal Field Theories with Global Symmetry,''
  J.\ Phys.\ A  {\bf 44}, 035402 (2011)
  [arXiv:1009.5985 [hep-th]].
}

\lref\LutyYE{
  M.~A.~Luty, T.~Okui,
  ``Conformal technicolor,''
JHEP {\bf 0609}, 070 (2006).
[hep-ph/0409274].
}

\lref\DineDV{
  M.~Dine, P.~J.~Fox, E.~Gorbatov, Y.~Shadmi, Y.~Shirman and S.~D.~Thomas,
  ``Visible effects of the hidden sector,''
  Phys.\ Rev.\  D {\bf 70}, 045023 (2004)
  [arXiv:hep-ph/0405159].
}

\lref\CohenQC{
  A.~G.~Cohen, T.~S.~Roy, M.~Schmaltz,
  ``Hidden sector renormalization of MSSM scalar masses,''
JHEP {\bf 0702}, 027 (2007).
[hep-ph/0612100].
}

\lref\RoyNZ{
  T.~S.~Roy, M.~Schmaltz,
  ``Hidden solution to the mu/Bmu problem in gauge mediation,''
Phys.\ Rev.\  {\bf D77}, 095008 (2008).
[arXiv:0708.3593 [hep-ph]].
}

\lref\PerezNG{
  G.~Perez, T.~S.~Roy, M.~Schmaltz,
  ``Phenomenology of SUSY with scalar sequestering,''
Phys.\ Rev.\  {\bf D79}, 095016 (2009).
[arXiv:0811.3206 [hep-ph]].
}

\lref\KaplanAC{
  D.~E.~Kaplan, G.~D.~Kribs and M.~Schmaltz,
  ``Supersymmetry breaking through transparent extra dimensions,''
  Phys.\ Rev.\  D {\bf 62}, 035010 (2000)
  [arXiv:hep-ph/9911293].
}

\lref\ChackoMI{
  Z.~Chacko, M.~A.~Luty, A.~E.~Nelson and E.~Ponton,
  ``Gaugino mediated supersymmetry breaking,''
  JHEP {\bf 0001}, 003 (2000)
  [arXiv:hep-ph/9911323].
}

\lref\SchmaltzGY{
  M.~Schmaltz and W.~Skiba,
  ``Minimal gaugino mediation,''
  Phys.\ Rev.\  D {\bf 62}, 095005 (2000)
  [arXiv:hep-ph/0001172].
}

\lref\DumitrescuHA{
  T.~T.~Dumitrescu, Z.~Komargodski, N.~Seiberg, D.~Shih,
  ``General Messenger Gauge Mediation,''
JHEP {\bf 1005}, 096 (2010).
[arXiv:1003.2661 [hep-ph]].
}

\lref\PolandWG{
  D.~Poland and D.~Simmons-Duffin,
  ``Bounds on 4D Conformal and Superconformal Field Theories,''
  arXiv:1009.2087 [hep-th].
}

\lref\WittenQJ{
  E.~Witten,
  ``Anti-de Sitter space and holography,''
  Adv.\ Theor.\ Math.\ Phys.\  {\bf 2}, 253 (1998)
  [arXiv:hep-th/9802150].
}

\lref\GubserBC{
  S.~S.~Gubser, I.~R.~Klebanov and A.~M.~Polyakov,
  ``Gauge theory correlators from non-critical string theory,''
  Phys.\ Lett.\  B {\bf 428}, 105 (1998)
  [arXiv:hep-th/9802109].
}

\lref\MaldacenaRE{
  J.~M.~Maldacena,
  ``The large N limit of superconformal field theories and supergravity,''
  Adv.\ Theor.\ Math.\ Phys.\  {\bf 2}, 231 (1998)
  [Int.\ J.\ Theor.\ Phys.\  {\bf 38}, 1113 (1999)]
  [arXiv:hep-th/9711200].
}

\lref\HeemskerkPN{
  I.~Heemskerk, J.~Penedones, J.~Polchinski and J.~Sully,
  ``Holography from Conformal Field Theory,''
  JHEP {\bf 0910}, 079 (2009)
  [arXiv:0907.0151 [hep-th]].
}

\lref\HeemskerkTY{
  I.~Heemskerk and J.~Sully,
  ``More Holography from Conformal Field Theory,''
  JHEP {\bf 1009}, 099 (2010)
  [arXiv:1006.0976 [hep-th]].
}

\lref\FitzpatrickZM{
  A.~L.~Fitzpatrick, E.~Katz, D.~Poland and D.~Simmons-Duffin,
  ``Effective Conformal Theory and the Flat-Space Limit of AdS,''
  arXiv:1007.2412 [hep-th].
}

\lref\DHokerJP{
  E.~D'Hoker, S.~D.~Mathur, A.~Matusis, L.~Rastelli,
  ``The Operator product expansion of N=4 SYM and the 4 point functions of supergravity,''
Nucl.\ Phys.\  {\bf B589}, 38-74 (2000).
[hep-th/9911222].
}

\lref\FreedmanBJ{
  D.~Z.~Freedman, S.~D.~Mathur, A.~Matusis, L.~Rastelli,
  ``Comments on 4 point functions in the CFT / AdS correspondence,''
Phys.\ Lett.\  {\bf B452}, 61-68 (1999).
[hep-th/9808006].
}

\lref\ArutyunovPY{
  G.~Arutyunov, S.~Frolov,
  ``Four point functions of lowest weight CPOs in N=4 SYM(4) in supergravity approximation,''
Phys.\ Rev.\  {\bf D62}, 064016 (2000).
[hep-th/0002170].
}

\lref\DolanTT{
  F.~A.~Dolan, H.~Osborn,
  ``Superconformal symmetry, correlation functions and the operator product expansion,''
Nucl.\ Phys.\  {\bf B629}, 3-73 (2002).
[hep-th/0112251].
}

\lref\AndrianopoliVR{
  L.~Andrianopoli, M.~Bertolini, A.~Ceresole, R.~D'Auria, S.~Ferrara and P.~Fre',
  ``General Matter Coupled N=2 Supergravity,''
  Nucl.\ Phys.\  B {\bf 476}, 397 (1996)
  [arXiv:hep-th/9603004].
}

\lref\AndrianopoliCM{
  L.~Andrianopoli, M.~Bertolini, A.~Ceresole, R.~D'Auria, S.~Ferrara, P.~Fre and T.~Magri,
  ``N = 2 supergravity and N = 2 super Yang-Mills theory on general scalar
  manifolds: Symplectic covariance, gaugings and the momentum map,''
  J.\ Geom.\ Phys.\  {\bf 23}, 111 (1997)
  [arXiv:hep-th/9605032].
}

\lref\CeresoleWI{
  A.~Ceresole, G.~Dall'Agata, R.~Kallosh and A.~Van Proeyen,
  ``Hypermultiplets, domain walls and supersymmetric attractors,''
  Phys.\ Rev.\  D {\bf 64}, 104006 (2001)
  [arXiv:hep-th/0104056].
}

\lref\deWitBK{
  B.~de Wit, M.~Rocek and S.~Vandoren,
  ``Gauging isometries on hyperKaehler cones and quaternion-Kaehler
  manifolds,''
  Phys.\ Lett.\  B {\bf 511}, 302 (2001)
  [arXiv:hep-th/0104215].
}

\lref\TachikawaTQ{
  Y.~Tachikawa,
  ``Five-dimensional supergravity dual of a-maximization,''
  Nucl.\ Phys.\  B {\bf 733}, 188 (2006)
  [arXiv:hep-th/0507057].
}

\lref\CeresoleZS{
  A.~Ceresole, G.~Dall'Agata, R.~D'Auria and S.~Ferrara,
  ``Spectrum of type IIB supergravity on AdS(5) x T(11): Predictions on N  = 1
  SCFT's,''
  Phys.\ Rev.\  D {\bf 61}, 066001 (2000)
  [arXiv:hep-th/9905226].
}

\lref\CeresoleHT{
  A.~Ceresole, G.~Dall'Agata and R.~D'Auria,
  ``KK spectroscopy of type IIB supergravity on AdS(5) x T(11),''
  JHEP {\bf 9911}, 009 (1999)
  [arXiv:hep-th/9907216].
}

\lref\DolanQI{
  F.~A.~Dolan and H.~Osborn,
  ``Applications of the Superconformal Index for Protected Operators and
  q-Hypergeometric Identities to N=1 Dual Theories,''
  Nucl.\ Phys.\  B {\bf 818}, 137 (2009)
  [arXiv:0801.4947 [hep-th]].
}

\lref\LutyJH{
  M.~A.~Luty and R.~Sundrum,
  ``Supersymmetry breaking and composite extra dimensions,''
  Phys.\ Rev.\  D {\bf 65}, 066004 (2002)
  [arXiv:hep-th/0105137].
}

\lref\SchmaltzQS{
  M.~Schmaltz and R.~Sundrum,
  ``Conformal sequestering simplified,''
  JHEP {\bf 0611}, 011 (2006)
  [arXiv:hep-th/0608051].
}

\lref\BarnesBM{
  E.~Barnes, E.~Gorbatov, K.~A.~Intriligator, M.~Sudano, J.~Wright,
  ``The Exact superconformal R-symmetry minimizes tau(RR),''
Nucl.\ Phys.\  {\bf B730}, 210-222 (2005).
[hep-th/0507137].
}

\lref\BreitenlohnerBM{
  P.~Breitenlohner and D.~Z.~Freedman,
  ``Positive Energy In Anti-De Sitter Backgrounds And Gauged Extended
  Supergravity,''
  Phys.\ Lett.\  B {\bf 115}, 197 (1982).
}

\lref\DolanUT{
  F.~A.~Dolan and H.~Osborn,
  ``Conformal four point functions and the operator product expansion,''
  Nucl.\ Phys.\  B {\bf 599}, 459 (2001)
  [arXiv:hep-th/0011040].
}

\lref\DolanHV{
  F.~A.~Dolan, H.~Osborn,
  ``Conformal partial waves and the operator product expansion,''
Nucl.\ Phys.\  {\bf B678}, 491-507 (2004).
[hep-th/0309180].
}

\lref\KlebanovTB{
  I.~R.~Klebanov, E.~Witten,
  ``AdS / CFT correspondence and symmetry breaking,''
Nucl.\ Phys.\  {\bf B556}, 89-114 (1999).
[hep-th/9905104].
}

\lref\HartmanDY{
  T.~Hartman, L.~Rastelli,
  ``Double-trace deformations, mixed boundary conditions and functional determinants in AdS/CFT,''
JHEP {\bf 0801}, 019 (2008).
[hep-th/0602106].
}

\lref\FreedmanTZ{
  D.~Z.~Freedman, S.~D.~Mathur, A.~Matusis and L.~Rastelli,
  ``Correlation functions in the CFT(d) / AdS(d+1) correspondence,''
  Nucl.\ Phys.\  B {\bf 546}, 96 (1999)
  [arXiv:hep-th/9804058].
}

\lref\DHokerGD{
  E.~D'Hoker, D.~Z.~Freedman,
  ``Gauge boson exchange in AdS(d+1),''
Nucl.\ Phys.\  {\bf B544}, 612-632 (1999).
[hep-th/9809179].
}

\lref\DHokerJC{
  E.~D'Hoker, D.~Z.~Freedman, S.~D.~Mathur, A.~Matusis, L.~Rastelli,
  ``Graviton and gauge boson propagators in AdS(d+1),''
Nucl.\ Phys.\  {\bf B562}, 330-352 (1999).
[hep-th/9902042].
}

\lref\DHokerNI{
  E.~D'Hoker, D.~Z.~Freedman and L.~Rastelli,
  ``AdS/CFT 4-point functions: How to succeed at z-integrals without  really
  trying,''
  Nucl.\ Phys.\  B {\bf 562}, 395 (1999)
  [arXiv:hep-th/9905049].
}

\lref\CecottiQN{
  S.~Cecotti, S.~Ferrara and L.~Girardello,
  ``Geometry of Type II Superstrings and the Moduli of Superconformal Field
  Theories,''
  Int.\ J.\ Mod.\ Phys.\  A {\bf 4}, 2475 (1989).
}

\lref\CassaniUW{
  D.~Cassani, G.~Dall'Agata and A.~F.~Faedo,
  ``Type IIB supergravity on squashed Sasaki-Einstein manifolds,''
  JHEP {\bf 1005}, 094 (2010)
  [arXiv:1003.4283 [hep-th]].
}

\lref\LiuSA{
  J.~T.~Liu, P.~Szepietowski and Z.~Zhao,
  ``Consistent massive truncations of IIB supergravity on Sasaki-Einstein
  manifolds,''
  Phys.\ Rev.\  D {\bf 81}, 124028 (2010)
  [arXiv:1003.5374 [hep-th]].
}

\lref\GauntlettVU{
  J.~P.~Gauntlett and O.~Varela,
  ``Universal Kaluza-Klein reductions of type IIB to N=4 supergravity in five
  dimensions,''
  JHEP {\bf 1006}, 081 (2010)
  [arXiv:1003.5642 [hep-th]].
}

\lref\SkenderisVZ{
  K.~Skenderis, M.~Taylor and D.~Tsimpis,
  ``A Consistent truncation of IIB supergravity on manifolds admitting a
  Sasaki-Einstein structure,''
  JHEP {\bf 1006}, 025 (2010)
  [arXiv:1003.5657 [hep-th]].
}

\lref\GennipThesis{
 Y.~v.~Gennip, ``Mathematical and physical aspects of the quaternion-K\"ahler manifold that arises in type IIA string compactification on rigid Calabi-Yau manifolds," Master's Thesis, http://www1.phys.uu.nl/wwwitf/Teaching/ouder/04YvanGennip.pdf
 }

\lref\AdamsSV{
  A.~Adams, N.~Arkani-Hamed, S.~Dubovsky, A.~Nicolis and R.~Rattazzi,
  ``Causality, analyticity and an IR obstruction to UV completion,''
  JHEP {\bf 0610}, 014 (2006)
  [arXiv:hep-th/0602178].
}


\noblackbox


\def\figin{\epsfcheck\figin}\def\figins{\epsfcheck\figins}
\def\epsfcheck{\ifx\epsfbox\UnDeFiNeD
\message{(NO epsf.tex, FIGURES WILL BE IGNORED)}
\gdef\figin##1{\vskip2in}\gdef\figins##1{\hskip.5in}
\else\message{(FIGURES WILL BE INCLUDED)}%
\gdef\figin##1{##1}\gdef\figins##1{##1}\fi}
\def\DefWarn#1{}
\def\figinsert{\goodbreak\midinsert}
\def\ifig#1#2#3{\DefWarn#1\xdef#1{fig.~\the\figno}
\writedef{#1\leftbracket fig.\noexpand~\the\figno}%
\figinsert\figin{\centerline{#3}}\medskip\centerline{\vbox{\baselineskip12pt
\advance\hsize by -1truein\noindent\footnotefont{\bf
Fig.~\the\figno:\ } \it#2}}
\bigskip\endinsert\global\advance\figno by1}

\newbox\tmpbox\setbox\tmpbox\hbox{\abstractfont }
\Title{\vbox{\baselineskip12pt }} {
\vbox{\centerline{
 Anomalous Dimensions of Non-Chiral  Operators   }
 \vskip 6pt
 \centerline{ from AdS/CFT} 
}}

\smallskip
\centerline{A. Liam Fitzpatrick$^1$ and David Shih$^2$}
\smallskip
\bigskip
\centerline{$^1${\it Dept.\ of Physics, Boston University, Boston, MA 02215 USA}} \centerline{$^2${\it 
NHETC, Dept.\ of Physics, Rutgers University, Piscataway, NJ 08854 USA}}
\vskip 1cm

\noindent  
Non-chiral operators with positive anomalous dimensions can have interesting applications to supersymmetric model building. Motivated by this, we develop a new method for obtaining the anomalous dimensions of non-chiral double-trace operators in $\CN=1$ superconformal field theories (SCFTs) with weakly-coupled AdS duals. Via the Hamiltonian formulation of AdS/CFT, we show how to directly compute the anomalous dimension as a bound state energy in the gravity dual. This simplifies previous approaches based on the four-point function and the OPE. We apply our method to a class of effective AdS$_5$ supergravity models, and we find that the binding energy can have either sign. If such models can be UV completed, they will provide the first calculable examples of SCFTs with positive anomalous dimensions.

\bigskip

\Date{April 2011}

\newsec{Introduction}

Recently, there has been significant progress in characterizing the operator product expansion (OPE) in 4d CFTs using general principles such as crossing symmetry and unitarity. For real scalar primary operators $\phi$, the authors of \refs{\RattazziPE,\RychkovIJ} were able to place an upper bound on the anomalous dimension of $\phi^2$,
the first scalar primary appearing in the $\phi \times \phi$ OPE. Here, following the literature, we define the anomalous dimension of $\phi^2$ to be:
\eqn\anomdimdef{
\gamma_{\phi^2} \equiv \Delta_{\phi^2} - 2\Delta_\phi .
}
In \RattazziYC, these bounds were extended to the case of a complex scalar field transforming under a global symmetry group.

Upper bounds on anomalous dimensions of composite operators in CFTs are interesting from a phenomenological point of view. In the context of non-supersymmetric CFTs, such bounds can have implications for conformal technicolor theories where the Higgs field $H$ is subject to strong conformal dynamics \LutyYE. In general, one wants $H^\dagger H$ to have dimension $\sim 4$ for the gauge hierarchy problem, but $H$ to have dimension $\sim 1$ for flavor. Thus one wants a large positive anomalous dimension for $H^\dagger H$. Upper bounds on anomalous dimensions  can constrain or rule out such technicolor models. 

In this paper, we will be interested in analogous issues in 4d superconformal field theories (SCFTs), where now we take $\phi$ to be a chiral primary operator. Here, interesting applications of positive anomalous dimensions arise when $\phi$ participates in SUSY-breaking.
Then the  soft masses in the MSSM come from operators of the form
\eqn\genops{
\int d^2 \theta\, {\phi \, \Phi_i \Phi_j \over M}, \ \ \ \ \
 \int d^4 \theta\,  { \phi^\dagger \phi\, \Phi_i \Phi_j \over M^2},
}
where $\Phi_i$ are MSSM  fields and $M$ is a UV scale where
these operators are generated. In general, the latter operators are unconstrained by supersymmetry, and in various contexts they can easily lead to too-large SUSY-breaking effects.  One way to suppress these effects is to imagine that  $\gamma_{\phi^\dagger\phi}>0$  due to strong SCFT dynamics. Then
 the coefficients of \genops\ run strongly as one flows into the IR, and at a scale $\mu\ll M$ one finds
\eqn\genopsrun{
\int d^2 \theta\, \left( {\mu \over M} \right)^{\Delta_\phi-1}{\phi \, \Phi_i \Phi_j \over M}, \ \ \ \ \
\int d^4 \theta\, \left( {\mu \over M} \right)^{2\Delta_{\phi^\dagger \phi}-2}{ \phi^\dagger \phi \,\Phi_i \Phi_j \over M^2}.
}
This mechanism is used, for example, in solutions to the $\mu/B_\mu$ problem  in gauge mediation \refs{\DineDV\CohenQC\RoyNZ-\PerezNG};  solutions to the problem
of flavor-violation in gravity mediation \refs{\KaplanAC\ChackoMI-\SchmaltzGY}; and gaugino mediation
in the context of ``general messenger gauge mediation" \DumitrescuHA.

Despite the many potential applications of positive anomalous dimensions, so far there do not actually exist any examples of SCFTs with $\gamma_{\phi^\dagger\phi}>0$. Thus it is interesting to speculate on whether $\gamma_{\phi^\dagger\phi}\le 0$ in all SCFTs. The authors of \PolandWG\ were able to prove a general upper bound on $\gamma_{\phi^\dagger\phi}$, by extending the crossing-symmetry-based methods of \RattazziPE\ to the supersymmetric case. Their bound allows for positive anomalous dimensions, but it can almost certainly be significantly improved with further numerical work, given the stronger results of \RattazziYC\ on complex scalars, which did not assume supersymmetry. Thus it is still conceivable that general SCFT constraints could imply the strongest possible bound $\gamma_{\phi^\dagger\phi}\le 0$.

In this paper, we will approach the question of positive anomalous dimensions in SCFTs in a complementary way. Rather than attempting to refine and improve the general bounds, we will instead use the tools of AdS/CFT to study explicit examples \refs{\MaldacenaRE\WittenQJ-\GubserBC}. As is well known, local supergravity theories in AdS$_5$ (supposing they have a stable UV
completion) provide constructive examples of a certain class of 4d SCFTs, namely theories which have a large $N$ 't Hooft limit.
SCFTs realized in this way have
conformally invariant correlation functions that satisfy the constraints
of crossing symmetry and unitarity, order by order in the $1/N$ expansion. Thus they are well-suited to exploring the space of possibilities consistent with  general bounds. 

As was recently emphasized in \refs{\HeemskerkPN\HeemskerkTY-\FitzpatrickZM}, an especially useful simplifying limit is where the AdS theory has only a handful of light states
(compared parametrically to the Planck scale), so that one may
decouple all massive string states and
focus only on a minimal light sector.   On the SCFT side, this corresponds to decoupling all but a small number of single-trace operators, and focusing on the multi-trace operators built out of these.

In the following sections, we will develop the necessary tools  for calculating $\gamma_{\phi^\dagger\phi}$ in such SCFTs. Since we are interested primarily in the sign of $\gamma_{\phi^\dagger\phi}$, we are free to focus on the leading-order effect in the $1/N$ expansion. This corresponds to doing semi-classical supergravity in an AdS$_5$ background. In section 2, we will describe the setup in more detail. For simplicity, we will focus on a single charged, complex scalar field $\phi$ minimally coupled to gravity and the graviphoton:
\eqn\fullactionintro{\eqalign{
S & = {1\over \kappa^2}\int d^5 x \sqrt{-g} \left[
- (D^\mu\phi)^\dagger\,(D_\mu\phi) 
 - m^2 \phi^\dagger\phi  -{a\over R^2} (\phi^\dagger \phi)^2- b (\phi^\dagger \phi)(\partial_\mu\phi^\dagger\partial^\mu\phi)\right]
}}
Here $a$ and $b$ are dimensionless coefficients that are a priori
free parameters of the model, and $R$ is the AdS radius.

Even after restricting to the leading-order effect in the $1/N$ expansion and to the minimal model \fullactionintro, calculations in AdS$_5$ can be dauntingly complex.  The conventional method  \refs{\DHokerJP\FreedmanBJ\ArutyunovPY-\DolanTT} for calculating the dimension of
$\phi^\dagger \phi$ in AdS/CFT has been to first obtain the four-point function 
$\langle\phi^\dagger(x_1)\phi(x_2)\phi^\dagger(x_3)\phi(x_4)\rangle$ using the standard techniques,   
go to a short-distance limit dominated by the OPE, and read off the anomalous 
dimensions from the expansion in conformal blocks. The correlation functions cannot be written in closed form,
but rather must be expressed in terms of special integral functions,
and the anomalous dimensions are related
to the coefficients of logarithmically-singular terms in these special functions.

In section 3, we will present a much simpler and more direct method for computing
anomalous dimensions. Our method is based on the Hamiltonian formulation of AdS/CFT, rather than the more commonly used Lagrangian formulation which leads to correlation functions.   
It makes use of the fact that the anomalous dimension of $\phi^\dagger\phi$ in the SCFT is dual to the binding energy of two-particle state in AdS. We show how to calculate the binding energies directly, thereby bypassing the four point function altogether. 

Our method also simplifies and extends previous approaches based on the Hamiltonian formalism, in particular the recent work of \FitzpatrickZM. We generalize the results of \FitzpatrickZM\ to include complex scalar fields with arbitrary gauge boson and graviton exchange. Furthermore, we show how all these different contributions to the binding energy can be understood in a uniform, semiclassical framework. Integrating out the gauge boson and graviton using their classical equations of motion, we will derive a non-local, quartic, effective interaction Hamiltonian $\delta H_{\rm eff}$ for the dual bulk field. This will allow us to obtain the binding energy of $\phi^\dagger\phi$ and $\phi \phi$ using {\it first-order} perturbation theory:
\eqn\gindinggeni{\eqalign{
\gamma_{\phi^\dagger\phi}  &= \int d^4x\,\sqrt{-g}\, \langle \phi^\dagger \phi | \delta H_{\rm eff}[\phi,\phi^\dagger] | \phi^\dagger \phi \rangle\cr
\gamma_{\phi\phi} &=  \int d^4x\,\sqrt{-g}\,\langle \phi \phi | \delta H_{\rm eff}[\phi,\phi^\dagger] | \phi \phi \rangle .}}
By contrast, in \FitzpatrickZM, it was necessary to go to {\it second-order} perturbation theory to deal with particle exchange. This in turn necessitated understanding the wavefunctions for an infinite tower of excited intermediate states, which was only carried out for $s$-channel exchange of scalar fields.

In general, the dimension of chiral operators is protected by supersymmetry,
so that $\gamma_{\phi\phi}$ necessarily vanishes.  Consequently, 
our calculation of $\gamma_{\phi \phi}$ imposes a 
relation on the coefficients $a$ and $b$ in \fullactionintro:
\eqn\abrelni{
a = \Delta(\Delta-2)b-{2\Delta^2\over3}.
}
Our final result for $\gamma_{\phi^\dagger \phi}$ turns out to be
\eqn\gammafinal{
\gamma_{\phi^\dagger \phi} \propto \left( b - {2\Delta(2\Delta+3)\over 3(2\Delta+1)}\right),
}
where the proportionality constant is strictly positive for $\Delta>1$. As a highly non-trivial check of our method, we rederive the $\phi\phi$ and $\phi^\dagger\phi$ anomalous dimensions using the conventional four-point function approach  in section 4. 

Interestingly, although \gammafinal\ is a binding energy between oppositely charged particles, it is not always negative. At large $\Delta$,
one can take a flat-space limit and $\gamma_{\phi^\dagger \phi}<0$ as expected. However, for $\Delta$ small, the AdS curvature
modifies the gravitational force and the anomalous dimension can have
either sign.

In section 5, we apply our general tools to the study of a specific, minimal 
supergravity theory for a single hypermultiplet, based on the coset $SU(2,1)/U(2)\times U(1)$. Starting from the explicit $\CN=2$ supergravity Lagrangian for the hypermultiplet sigma model, we derive the Lagrangian \fullactionintro\ with specific values of $a$ and $b$. We find that for a range of choices of the parameters of this model, one has $\gamma_{\phi^\dagger \phi} >0$. Finally, in section 6, we conclude with a summary of our results, and a discussion of potential model-building applications and future directions.

\newsec{General scalar fields coupled to $\CN=2$ supergravity}

\subsec{Brief overview of $\CN=2$ $d=5$ supergravity}

In this section, we will briefly describe the structure of AdS$_5$ theories which are dual to 4d $\CN=1$ SCFTs in the large $N$ limit at strong 't Hooft coupling. At low energies and weak coupling, the effective AdS$_5$ theory falls into the framework of $\CN=2$, $d=5$ supergravity. Useful references include  \refs{\AndrianopoliVR\AndrianopoliCM\CeresoleWI\deWitBK-\TachikawaTQ}. 

Chiral primaries in the SCFT are dual to hypermultiplets in the 5d bulk. A hypermultiplet has four real degrees of freedom; these are dual to the chiral primary and its $F$-component.

Besides the hypermultiplet, the other basic BPS multiplets in $\CN=2$, $d=5$ supergravity are the vector multiplet and the gravity multiplet. The gravity multiplet is dual to the multiplet containing the stress tensor and the $U(1)_R$-current of $\CN=1$ SCFT. Its bosonic degrees of freedom consist of the funfbein and the graviphoton. Meanwhile, vector multiplets are dual to the current supermultiplets of global symmetries in the SCFT. Each vector multiplet's bosonic degrees of freedom consist of a real scalar and a gauge field. 

Supersymmetry dictates that the vector multiplet scalars take values on a ``special K\"ahler manifold" and the hypermultiplet scalars take values on a ``quaternion K\"ahler manifold." Interactions arise from gauging isometries in these manifolds.

The gravity and vector multiplet are examples of ``massless multiplets," and the hypermultiplets are examples of ``chiral multiplets." These multiplets are protected by supersymmetry and are dual to shortened representations of the $\CN=1$ superconformal algebra in $d=4$.  In addition, there are many other possible multiplets in $\CN=2$, $d=5$ supergravity. These include higher-spin shortened multiplets. Others, known as ``massive multiplets," are less constrained by supersymmetry, and are dual to ``semi-short" and ``long" representations of the $\CN=1$ superconformal algebra in $d=4$. These are described in detail in \refs{\CeresoleZS,\CeresoleHT}. (See also the appendix of  \DolanQI\ for a concise summary of the unitary representations of the $\CN=1$ superconformal algebra.) Massive multiplets generally correspond to KK modes, so they will usually be present in any compactification down to 5d. 

We note that while the representations of the $\CN=1$ superconformal algebra include multiplets with arbitrarily high spin,  
fields with spin $>2$ are not believed to arise in any local, weakly-coupled supergravity theory. So we will assume these are absent in the effective AdS theory, and  focus our attention on spin $\le 2$.

In addition, we will restrict ourselves  to supergravity theories without massless vector multiplets, for the following reason. 
As discussed above, massless vector multiplets in AdS are dual to conserved non-R global symmetries in the SCFT. As emphasized by many authors (see e.g.\ \refs{\LutyJH,\DineDV,\SchmaltzQS,\RoyNZ}), theories where $\phi$ is charged under a (non-R) global symmetry necessarily have $\gamma_{\phi^\dagger\phi}\le 0$.\foot{In such theories, the conserved current of the global symmetry lives in a protected multiplet whose lowest component is a dimension-2 scalar $J$, and $J$ must appear in the OPE of
$\phi^\dagger \times \phi$ with a nonzero coefficient fixed by the global symmetry. Since unitarity restricts $\Delta_\phi\ge 1$, one necessarily has $\gamma_{\phi^\dagger\phi} \le \Delta_J - 2 \Delta_\phi\le 0$.} Given that we are interested in the possibility of $\gamma_{\phi^\dagger\phi}> 0$, it makes sense to exclude massless vector multiplets from our setup.

So, to summarize, for the problem we are interested in, we can consider hypermultiplets coupled to themselves, to the gravity multiplet, and to other massive multiplets with spin $\le 2$. Now let us describe the setup in more detail.

\subsec{Our setup}

We are interested in obtaining the anomalous dimension of $\phi^\dagger\phi$ to leading order in the $1/N$ expansion.  According to the AdS/CFT dictionary, $1/N$ corresponds to the 5d gravitational coupling $\kappa = \sqrt{8\pi G_N}$ in the bulk dual
\eqn\Nreln{
{1\over N} \sim {1\over ( m_5 R)^{3/2}}  \sim { \kappa \over R^{3/2} },
}
where $m_5$ is the five-dimensional Planck scale and $R$ is the AdS radius. In what follows, we will work in units of $R=1$. Then $\kappa$ controls the strength of the interactions in the 5d supergravity theory. To compute the leading anomalous dimension, we must specify the interactions of $\phi$ with itself and with all the other fields in the 5d bulk dual, to lowest order in $\kappa$. As we will see below, the leading anomalous dimensions arise at $\CO(\kappa^2)$ and come from tree-level diagrams in the bulk theory.

As discussed above, $\phi$ is dual to a complex scalar field (which we will denote with the same symbol) inside a hypermultiplet. Clearly, $\phi$ must couple canonically to gravity and to the graviphoton. It can also couple to itself via quartic interactions. (Cubic self-interactions are forbidden by charge conservation.) So far we have described the following setup:
\eqn\fullaction{
S = {1\over \kappa^2}\int d^5 x \sqrt{-g} \left(   {1 \over 2}
\left({\cal R} + 12 \right) -{1 \over 4g^2 } F^2 - (D^\mu\phi)^\dagger\,(D_\mu\phi) 
 - m^2 \phi^\dagger\phi  - V[\phi,\phi^\dagger] \right) .
}
where $D_\mu = \partial_\mu - i q A_\mu$ and 
\eqn\scalarpot{\eqalign{
V[\phi,\phi^\dagger] & = a (\phi^\dagger \phi)^2 + b (\phi^\dagger \phi)
 (\partial_\mu \phi^\dagger \partial^\mu \phi) 
}}
for some coefficients $a$ and $b$.  Note that in \fullaction, all the fields are dimensionless and the only dependence on $\kappa$ is out front. Thus all the dimensions in \fullaction\ are made up with powers of $R$. 

Demanding the canonical 
relation between the central charge of the energy-momentum tensor
and the central charge of the R-current fixes\foot{
Conserved currents $J^\mu$ and the energy-momentum tensor $T^{\mu\nu}$ 
generate symmetry transformations in the CFT,
which provides them a canonical normalization based on their three-point
funtions with other operators.  
The central charges $c_V$ and $c_T$ of $J^\mu$ and $T^{\mu\nu}$ respectively
are then defined through their two-point functions:
\eqn\JTtwopoint{
\langle J^\mu (x) J^\nu(0) \rangle = {12 c_V \over (2\pi)^4} {I^{\mu\nu} \over x^6}, \ \ \ \ \langle T^{\mu\nu}(x) T^{\rho \sigma}(0) \rangle = 
{40 c_T \over \pi^4} \left( {1 \over 2}(I^{\mu\rho} I^{\nu\sigma} + I^{\mu\sigma} I^{\nu \rho}) - {1 \over 4} \delta^{\mu\nu} \delta^{\rho \sigma}  \right) ,}
where
$I^{\mu\nu} = \delta^{\mu\nu} - 2 {x^\mu x^\nu \over x^2}$ .
From the action \fullaction\ for $A_\mu$ and $h_{\mu\nu}$, 
one may calculate the two-point functions of $J^\mu$ and $T^{\mu\nu}$ and
compare to \JTtwopoint, with the result
$c_V = {8 \pi^2 \over g^2 \kappa^2 }, c_T = {\pi^2 \over \kappa^2}$.
However, since $J^\mu$ here is the R-current, it falls in the same multiplet 
as $T^{\mu\nu}$. Consequently, the central charges are related by
$c_V = {16 \over 3} c_T$ \BarnesBM, which enforces the
value for $g$ stated in the text.
}  
\eqn\gfixed{
g^2 = {3 \over 2}
}
According to the usual AdS/CFT dictionary, the scalar mass-squared is related to the dimension in the SCFT via:
\eqn\mDeltarel{
m^2 = \Delta(\Delta-4).
} 
The $U(1)$ charge here is not independent of the mass, but is given by 
\eqn\susyq{
q = {2 \over 3 } \Delta.
}
This relation is the bulk counterpart of the usual dimension/R-charge relation for chiral primaries in $\CN=1$ SCFT. 
Finally, $a$ and $b$ are not independent in this class of models. Since $\phi$ is a chiral primary in the SCFT, the dimension of $\phi^2$ is protected. 
Therefore, the energy of the $\phi\phi$ two-particle state must be exactly $2\Delta$.  As we will see in section 3.1, this imposes a relation between $a$ and $b$:
\eqn\abreln{
a = \Delta(\Delta-2)b-{2\Delta^2\over3}.
}
We believe this is a new relation between the quartic hypermultiplet couplings of $\CN=2$, $d=5$ supergravity theories. Below in section 5, we will see that it is respected in the specific example of the ``universal hypermultiplet." It would be interesting to test this relation further in more general examples and in actual string compactifications.

The action \fullaction, subject to the relations above, is the most general setup we will consider in this paper. According to the discussion in the previous subsection, we are ignoring couplings to additional massive multiplets. This is purely for simplicity; our methods should be easily extendable to include these modes as well, and it would be very interesting to do so. 

Aside from the possibility of massive multiplets, we claim that  \fullaction\ can be used to calculate the binding energies of $\phi\phi$ and $\phi^\dagger\phi$ in any $\CN=2$, $d=5$ supergravity theory, to leading order in the gravitational coupling $\kappa$. To prove this claim, it is useful to keep in mind some diagrammatic intuition. As we will describe in more detail in section 4, the binding energies arise from diagrams with four $\phi$ (or $\phi^\dagger$) external legs. Then the leading order binding energies arise at $\CO(\kappa^2)$ from single insertions of the quartic potential \scalarpot, as well as from tree-level graviton and gravi-photon exchange. All other supergravity interactions (higher order self-interactions of $\phi$, interactions with fermions and other charged scalars) can contribute only at loop order. Thus they involve more internal lines, and hence more powers of $\kappa$. We can also consider higher derivative corrections due to massive string states. These are suppressed by powers of $\alpha'$ inside the parentheses in \fullaction.
Since we are assuming $\alpha' \ll R^2$, these are also subleading effects. 
We conclude that \fullaction\ captures the leading-order binding energies of $\phi\phi$ and $\phi^\dagger\phi$, up to possible couplings to additional massive multiplets. 

Finally, we should emphasize that the effective theory \fullaction\ is  not UV complete. Thus any results derived from it are subject to the usual caveats of whether a given effective field theory can be UV-completed. In
practice this is usually accomplished by finding a string theory embedding. This extremely interesting line of investigation is beyond the scope of this paper; we look forward to returning to it in a future publication.

The form of the action \fullaction\ is ideally suited for power-counting
of $\kappa$. However, for computations, it is more convenient to canonically
normalize the fields 
\eqn\cannormfields{
(\phi, A_\mu, h_{\mu\nu}) \rightarrow \kappa
(\phi,g A_\mu, h_{\mu\nu})
} 
which we shall do in the following sections.

\newsec{A new approach to anomalous dimensions in AdS/CFT}

\subsec{Hamiltonian formulation of AdS/CFT -- the free theory}

In this section, we will derive general formulas for the leading order binding energies of $\phi\phi$ and $\phi^\dagger\phi$ in the Hamiltonian formulation of AdS/CFT. By directly focusing from the outset on the Hamiltonian and the spectrum of operators, we can bypass the
correlation functions of the theory and all the extra complications they bring. 

Our strategy will be to integrate out the photon and graviton in \fullaction, leaving behind an effective action for the scalar alone. Passing from the effective action to the Hamiltonian, the leading-order binding energy can be read off using first-order perturbation theory,
simply by looking at matrix elements of the Hamiltonian.  As we will see, most of the
effort of calculating the binding energies in the present method will
go into integrating out the photon and graviton.  Even this step will
be performed in a very physically transparent way, essentially by
treating them as semi-classical fields sourced by the two-particle
states, $\phi^\dagger \phi$ or $\phi \phi$, whose binding energies we wish
to calculate.  The high degree of symmetry of these sources 
drastically simplifies the response of the photon and graviton fields,
which can be solved for in simple, closed form for any value of $\Delta$.

To begin, let us describe the Hamiltonian formulation of the free theory ($\kappa\to 0$). We work with the Hamiltonian which generates time evolution in AdS$_5$ global coordinates: 
\eqn\metricconv{
ds^2 = {1\over \cos^2 \rho} \left( - dt^2 + d \rho^2 + \sin^2 \rho\,
d\Omega_3^2 \right).
}
(So $\rho\in [0,{\pi\over2})$ with the boundary of AdS occurring at $\rho\to\pi/2$.) As is well known, this Hamiltonian corresponds to that of the dual CFT in radial quantization, i.e.\ the dilatation operator. Thus its spectrum should correspond to the dimension of operators in the dual CFT. Indeed, for a free scalar field in AdS, the spacetime curvature
acts like a potential well and the energy spectrum of single particle $\phi$ states is discrete:
\eqn\especfree{
E_{n,\ell}^{(0)} = \Delta + 2n +\ell
}
where $\ell$ is the total spin.  This is  
in direct correspondence with the dimensions
of a single-trace scalar primary operator $\phi$  and its descendants in the dual
CFT \refs{\WittenQJ,\BreitenlohnerBM}. More generally, 
all of the single-particle states of AdS are in one-to-one correspondence
with the single-trace states in the CFT in radial quantization. 
In AdS, as in flat space, the creation and annihilation operators
for these states and their anti-particles are used to construct the 
second-quantized scalar field $\phi(x)$:
\eqn\phiexpand{
\phi(x) = \sum_{n,l,J} ( \psi_{n,l,J}^*(x)a_{n,l,J} + \psi_{n,l,J}(x) b_{n,l,J}^\dagger)
}
The sum is over the discrete eigenmodes of AdS, with $J$ labelling additional
spin quantum numbers (e.g. azimuthal spin).
$\psi_{n,l,J}$ are the
appropriate Klein-Gordon wavefunctions,  normalized so that the Hamiltonian constructed
from the conventional free scalar Lagrangian is simply
\eqn\freeH{
H_{\rm free} = \sum_{n,l,J} E^{(0)}_{n,l} \left( a_{n,l,J}^\dagger a_{n,l,J}
+ b_{n,l,J}^\dagger b_{n,l,J} \right) .
}
The only wavefunction that will turn out to be relevant to our calculation is that
of the lowest-energy mode:
\eqn\groundpsi{
\psi_0 (x) \equiv \psi_{0,0,0}(x) = N_\Delta (e^{i t} \cos\rho)^\Delta , \ \ \ \ \ N_\Delta = 
\sqrt{\Delta-1 \over  2\pi^2} .
}
Here the $2\pi^2$ in the denominator is the volume of $S^3$. The relevant two-particle states in the free theory are just
\eqn\phicreann{
|\phi\phi \rangle = {1\over\sqrt{2}}b_0^\dagger b_0^\dagger|0\rangle, 
\ \ \ \ \ |\phi^\dagger\phi\rangle = a_0^\dagger b_0^\dagger |0\rangle,
}
($a_0\equiv a_{0,0,0}, b_0 \equiv b_{0,0,0}$).
In the free theory, these states clearly have energy that is just
$2\Delta$, as one can verify by acting on them with $H_{\rm free}$.   

\subsec{Adding interactions}

Now consider turning on $\kappa$-suppressed interactions in \fullaction\ (remembering that we have canonicalized the fields via \cannormfields). These deform the Hamiltonian away from \freeH, 
\eqn\Hdeform{
H_{\rm free} \rightarrow H_{\rm free} +\kappa\, \delta H_{\rm exchange}+\kappa^2\, \delta H_{\rm contact}
} 
Here we have separated out the interactions due to photon and graviton exchange, which start at $\CO(\kappa)$, and the quartic scalar contact interactions, which start at $\CO(\kappa^2)$. We would like to find the perturbed spectrum of two particle states. One idea would be to directly apply time-independent perturbation theory to \Hdeform. Then the energy of $\phi \phi$ would be given by \FitzpatrickZM:
\eqn\tipt{
E_{\phi\phi} = E_{\phi\phi}^{(0)} +\kappa^2 \left( \< \phi \phi | \delta H_{\rm contact} | \phi \phi \>
  + \sum_{|\alpha\> \ne |\phi\phi\>} { | \< \phi \phi | \delta H_{\rm exchange} | \alpha \>|^2
\over E_{\phi\phi}^{(0)} - E_\alpha^{(0)} } \right) + \CO(\kappa^4), 
}
with an analogous formula for $E_{\phi^\dagger\phi}$.  So to extract the effect of photon and graviton exchange using \tipt, we must perform second order perturbation theory. This is a formidable task -- the sum in \tipt\ is over all possible intermediate states in the Hilbert space, including descendants and states with arbitrary spin.  
We will now instead develop a better approach that allows us to treat
all contributions using first-order perturbation theory.

As described at the beginning of this section, our approach is to integrate out the photon and graviton classically, leaving behind a non-local effective potential for the scalar field alone. Formally, we can write the effective theory for $\phi$ as:
\eqn\reducedaction{\eqalign{
S_{\rm eff} & = \int d^5 x \sqrt{-g} \left( - \left( |\partial \phi|^2
 + m^2 |\phi|^2 \right)  - V_{\rm eff}[\phi,\phi^\dagger] +\CO(\phi^6)\right) , \cr
V_{\rm eff}[\phi,\phi^\dagger] & = V[\phi,\phi^\dagger] - {1 \over 2} \kappa\, A_\mu[\phi, \phi^\dagger] J^\mu[\phi, \phi^\dagger] -{1 \over 4}\kappa \, h_{\mu\nu}[\phi,\phi^\dagger]
T^{\mu\nu}[\phi, \phi^\dagger]  \cr
}}
In this formula, $A_\mu[\phi,\phi^\dagger]$ and $h^{\mu\nu}[\phi,\phi^\dagger]$ are understood to be non-local functionals
of $\phi$ given by the solutions to the usual (linearized)  field equations for electromagnetism and gravity in curved space. We can write these
\eqn\photgraveom{\eqalign{
&\Delta_V^{\mu\nu} A_\nu[\phi,\phi^\dagger] = - \kappa\, J^\mu[\phi,\phi^\dagger]\cr
& \Delta_T^{\mu\nu,\rho\sigma} h_{\rho\sigma}[\phi,\phi^\dagger] = -{1\over2} \kappa\, T^{\mu\nu}[\phi,\phi^\dagger]\cr
}}
where $\Delta_V$ and $\Delta_T$ are second-order differential operators in the AdS coordinates, and $J_\mu$ and $T_{\mu\nu}$ are the usual $U(1)$ current and the stress tensor operators, respectively:
\eqn\JTdef{
J_\mu = i g q (\phi \partial_\mu \phi^\dagger - \phi^\dagger \partial_\mu \phi ),\qquad T_{\mu\nu} = (\partial_{\mu} \phi \partial_{\nu} \phi^\dagger +(\mu\leftrightarrow \nu))- g_{\mu\nu}
(\partial_\rho \phi \partial^\rho \phi^\dagger + m^2 \phi \phi^\dagger )
}
Of course, to even define $A_\mu[\phi, \phi^\dagger]$ and 
$h^{\mu\nu}[\phi, \phi^\dagger]$ via \photgraveom, it is necessary to choose a gauge. We will do this below; the final result for the anomalous dimensions is independent of this choice.

According to \photgraveom, $V_{\rm eff}$ is formally quartic in $\phi$, $\phi^\dagger$ and is $\CO(\kappa^2)$. All other terms in the Lagrangian (e.g.\ $A_\mu A^\mu \phi^\dagger \phi$) contribute to the effective potential only at higher order in $\phi$, and hence higher order in $\kappa$. 
Performing canonical quantization,  we find that the leading order interaction Hamiltonian density  
$\delta H_{\rm eff}$ is also just $V_{\rm eff}$.\foot{
From \reducedaction, we compute the conjugate momenta
\eqn\conjmomenta{
  \Pi_\phi = {\partial\CL\over \partial\dot\phi} = \dot\phi^\dagger - {\partial V_{\rm eff}\over\partial\dot\phi}, \ \ \ \ \ \
 \Pi_{\phi^\dagger} = {\partial\CL\over \partial\dot\phi^\dagger} = \dot\phi -  {\partial V_{\rm eff}\over\partial\dot\phi^\dagger},
 }
and then the Hamiltonian density:
\eqn\reducedhamil{\eqalign{
 \CH = \Pi_\phi\dot\phi+\Pi_{\phi^\dagger}\dot\phi^\dagger - \CL & = |\Pi_\phi|^2 + |\grad \phi|^2+ m^2|\phi|^2+  V_{\rm eff} +\dots = \CH_{\rm free} +  V_{\rm eff} + \dots
}}
where $\dots$ is higher order in $\phi$ or $\kappa$. Note that the linear terms $\propto {\partial V_{\rm eff}\over\partial\dot\phi}$ cancel out when passing to the fields and their conjugate momenta.
} Therefore, we can use first order perturbation theory to obtain the leading $\CO(\kappa^2)$ binding energies of the states $|\phi \phi\rangle$ and $| \phi^\dagger \phi \rangle$:\foot{Here we ignore self-contractions inside $V_{\rm eff}$, i.e.\ we treat it as normal ordered. Such self-contractions correct the mass of $\phi$ itself, and cancel out of the leading-order binding energy.}
\eqn\bindinggen{\eqalign{
& \gamma_{\phi\phi} =  \int d^4x\,\sqrt{-g}\,\langle \phi \phi | V_{\rm eff}[\phi,\phi^\dagger] | \phi \phi \rangle = 2\int d^4x\,\sqrt{-g}\, V_{\rm eff}[\phi,\phi^\dagger]\big|_{b_0^\dagger b_0^\dagger b_0 b_0}\cr
& \gamma_{\phi^\dagger\phi}  = \int d^4x\,\sqrt{-g}\, \langle \phi^\dagger \phi | V_{\rm eff}[\phi,\phi^\dagger] | \phi^\dagger \phi \rangle  =
\int d^4x\,\sqrt{-g}\, V_{\rm eff}[\phi,\phi^\dagger]\big|_{a_0^\dagger b_0^\dagger a_0 b_0}
}}  
This is the main result of this paper. As we shall see, it represents a significantly simpler and more direct calculation of the anomalous dimensions in AdS/CFT compared with previous approaches based on the four point function. In essence, the simplification gained here is due to the fact that the four point function requires full knowledge of the photon and graviton propagators in AdS$_5$, while this approach only requires certain matrix elements \bindinggen\ of the propagators. 

The only nontrivial step at this point is solving for the matrix elements of \photgraveom, and even here we will find astonishing simplifications due to the high degree of symmetry of the source wavefunctions \groundpsi. In the following subsections, we will flesh out the rest of this calculation. Some miscellaneous technical details are relegated to appendix A. Since $V_{\rm eff}$ is proportional to $\kappa^2$, we will set $\kappa=1$ below to avoid cluttering the equations. 

\subsec{Calculation of $\phi\phi$ anomalous dimension}

Let us first evaluate $\gamma_{\phi\phi}$, which is simpler. According to \bindinggen, we need to extract $b_0^\dagger b_0^\dagger b_0 b_0$ from $V_{\rm eff}[\phi,\phi^\dagger]$. Since $V_{\rm eff}$ is quartic in $\phi$, $\phi^\dagger$, there is only one way to do this given the expansion \phiexpand, namely to pull $\psi_0(x) b_0^\dagger$ from $\phi$ and $\psi_0^*(x) b_0$ from $\phi^\dagger$. So \bindinggen\ becomes
\eqn\phiphi{\eqalign{
\gamma_{\phi\phi} =2\int d^4x\,\sqrt{-g}\, V_{\rm eff}[\phi(x) = \psi_0(x),\phi^\dagger(x) = \psi_0^*(x)]\cr
}}
Using \reducedaction, the individual quartic, photon, and graviton contributions to \phiphi\ are as follows:

\item{1.} The easiest term in $V_{\rm eff}$ to evaluate is the contribution from
the original scalar potential, $V$, since this simply involves substituting
$\psi_0(x)$, $\psi_0^*(x)$ from \groundpsi\ in the appropriate places.  Performing this, we obtain
\eqn\quarticgamma{
\gamma_{\phi \phi}^{(quartic)} =  { \pi^2 N_\Delta^4(a + b \Delta (2 - \Delta))\over  (\Delta-1)(2\Delta-1)}
 .}

\item{2.} Next, consider the photon contribution.  According to \reducedaction\ and \phiphi, this takes the form
\eqn\phiphiphoton{
\gamma_{\phi\phi}^{(photon)} = -\int d^4x\,\sqrt{-g}\,A_\mu[\psi_0(x),\psi_0^*(x)]J^\mu[\psi_0(x),\psi_0^*(x)]
}
where now $A_\mu$ is now a function of $x$ (rather than an operator) which satisfies \photgraveom\ with $\phi(x)\to \psi_0(x)$, $\phi^\dagger(x)\to \psi_0^*(x)$. We note that $J^\mu[\psi_0(x),\psi_0^*(x)]$ is time-independent, and has a very simple form: 
\eqn\chargedensityII{\eqalign{
J^0[\psi_0(x),\psi_0^*(x)] = - 2\Delta N_\Delta^2  g q \,    y^{2\Delta +2} , \ \ \ \ \  J^i[\psi_0(x),\psi_0^*(x)] =0.
}}
where $y\equiv \cos\rho$. The lack of any current component or time-dependence means that we may choose a gauge where the
only non-vanishing component of $A_\mu$ is the potential $A_0$.  Solving \photgraveom\ with the source \chargedensityII, we find a simple formula for $A_0$ (for details, see appendix A):
\eqn\potlfinsoln{
A_0 = - { N_\Delta^2 g q   ( y^2-y^{2\Delta})\over 2 (\Delta-1)(1-y^2)}\,.
}
Substituting back into \phiphiphoton, we obtain
\eqn\phiphigammaphot{
\gamma_{\phi \phi}^{(photon)} = {\pi^2 N_\Delta^4 g^2 q^2  \over 2\Delta-1} .
}

\item{3.} Finally, let us evaluate the graviton contribution.  According to \reducedaction\ and \phiphi, this takes the form
\eqn\phiphigrav{
\gamma_{\phi\phi}^{(graviton)} = -{1\over2}\int d^4x\,\sqrt{-g}\,h^{\mu\nu}[\psi_0(x),\psi_0^*(x)]T_{\mu\nu}[\psi_0(x),\psi_0^*(x)]
}
The energy-momentum tensor $T_{\mu\nu} [\psi_0,\psi_0^*]$ is again time-independent,
with a simple form:
\eqn\enmomII{
T^\mu_{\ \nu}[\psi_0(x),\psi_0^*(x)]  = 2\Delta N_\Delta^2 \, y^{2\Delta}  \cdot  {\rm diag} \left( 2-\Delta,
2, 2-\Delta+\Delta\, y^2, \dots, 2-\Delta+\Delta\, y^2 \right).
}
Solving \photgraveom\ with the source \enmomII, we again find a simple result:
\eqn\fgfinalII{\eqalign{
h_{tt} & = {2\Delta N^2_\Delta  (y^2-y^{2\Delta} ) \over 3 (\Delta -1)(1-y^2)},\qquad h_{\rho\rho}  = h_{tt} - {2\Delta N_\Delta^2 \over 3} y^{2\Delta -2}.
}}
with all other metric components vanishing. The energy shift for $\phi \phi$ 
may now be calculated by substituting \enmomII\ and \fgfinalII\ into 
 \phiphigrav. We obtain
\eqn\VgravI{
\gamma_{\phi\phi}^{(graviton)} =  -{2\pi^2 N_\Delta^4\Delta^2 (\Delta-2) \over 3 
(\Delta-1)(2\Delta-1)} .
}

\bigskip

Finally, let us put together all the different contributions. The $\phi\phi$ anomalous dimension is the sum of all contributions \quarticgamma, \phiphigammaphot, and \VgravI. Because  $\phi\phi$ is a chiral operator,
its dimension is protected by supersymmetry, so $\gamma_{\phi\phi}$ must vanish. Using \gfixed\ and \susyq,
the result is
\eqn\phiphigammatot{
\gamma_{\phi\phi} = \gamma_{\phi\phi}^{(quartic)} + \gamma_{\phi\phi}^{(photon)}
+\gamma_{\phi\phi}^{(graviton)} =  {\pi^2 N_\Delta^4 (2\Delta^2 +3 a -3 b \Delta (\Delta-2)) 
  \over 3 (\Delta-1)(2\Delta-1)}  = 0. 
}
where the fact that $\gamma_{\phi\phi}$ must vanish enforces the relation \abreln\ between $a$ and $b$.

\subsec{Calculation of $\phi^\dagger\phi$ anomalous dimension}

The calculation of $\gamma_{\phi^\dagger\phi}$ in \bindinggen\ is more difficult, because there are multiple ways to get $a_0^\dagger b_0^\dagger a_0 b_0$ from $V_{\rm eff}[\phi,\phi^\dagger]$. Essentially this is because now both $t$-channel and $s$-channel terms contribute.  To aid in handling all possible contractions systematically, let us
label each of the four $\phi$'s in $V_{\rm eff}$ separately:
\eqn\tildeVcontract{\eqalign{
V_{\rm eff}[\phi_1, \phi_2; \phi_1^\dagger, \phi_2^\dagger] & =
a (\phi_1 \phi_1^\dagger)(\phi_2 \phi_2^\dagger) +b(\phi_1 \phi_1^\dagger)(\partial_\mu \phi_2 \partial^\mu
\phi_2^\dagger) \cr
&- {1 \over 2}A_\mu[ \phi_1,\phi_1^\dagger] J^\mu[ \phi_2,\phi_2^\dagger] - {1 \over 4}
h^{\mu\nu}[ \phi_1,\phi_1^\dagger] T_{\mu\nu}[ \phi_2,\phi_2^\dagger] ,
}}
Thus we have $V_{\rm eff}[\phi,\phi^\dagger] = V_{\rm eff}[\phi_1=\phi,\phi_2=\phi;\phi_1^\dagger=\phi^\dagger,\phi_2^\dagger=\phi^\dagger]$. 

Note that there is a symmetry of $V_{\rm eff}$ under interchange of the
$\phi_i$'s that will be useful
in reducing the number of terms to be evaluated. If we switch
$\phi_1 \leftrightarrow \phi_2$ and $\phi_1^\dagger \leftrightarrow \phi_2^\dagger$,
then via integration by parts $V_{\rm eff}$ remains unchanged:
\eqn\Vsym{
V_{\rm eff}[\phi_1,\phi_2;\phi_1^\dagger,\phi_2^\dagger] = V_{\rm eff}[\phi_2,\phi_1;\phi_2^\dagger,\phi_1^\dagger] .
}
This identity is true term by term in \tildeVcontract.

With this new notation, the contribution from all contractions is very simply stated. Extracting $a_0^\dagger b_0^\dagger a_0 b_0$ from \tildeVcontract, and using \Vsym, we obtain
\eqn\phistarphi{\eqalign{
\gamma_{\phi^\dagger\phi} &=2 \int d^4x\,\sqrt{-g}\, V_{\rm eff}[\phi_1 = \psi_0(x),\phi_2 = \psi_0^*(x);\phi_1^\dagger= \psi_0(x),\phi_2^\dagger= \psi_0^*(x)]\cr
&\qquad+2\int d^4x\,\sqrt{-g}\, V_{\rm eff}[\phi_1 = \psi_0(x),\phi_2 = \psi_0^*(x);\phi_1^\dagger= \psi_0^*(x),\phi_2^\dagger= \psi_0(x)]
}}
The individual quartic, photon and graviton contributions are now (keep in mind we are setting $\kappa=1$ for convenience):

\item{1.} The easiest term in $V_{\rm eff}$ to evaluate is again the contribution from
the original scalar potential, $V$, since this simply involves substituting
$\psi_0(x)$, $\psi_0^*(x)$ in the appropriate places.  Performing this, we obtain
\eqn\quarticgammaphicphi{
\gamma_{\phi^\dagger \phi}^{(quartic)} = {2\pi^2 N_\Delta^4(a+b \Delta)   \over
(\Delta-1)(2\Delta-1)} \, .
}

\item{2.} Next, consider the photon contribution from \Vsym.  This is, explicitly,
\eqn\gammaphotphicphi{
\gamma_{\phi^\dagger\phi}^{(photon)} =-\int d^4x\,\sqrt{-g}\, \Big( A_\mu[\psi_0,\psi_0]J^\mu[\psi_0^*,\psi_0^*] + A_\mu[\psi_0,\psi_0^*]J^\mu[\psi_0^*,\psi_0]\Big)
}
Since $J^\mu[\psi_1,\psi_2]$ is antisymmetric in its arguments, the first term clearly vanishes, and the second gives the opposite of the $\phi\phi$ anomalous dimension \phiphiphoton. Therefore we conclude that
\eqn\phicphigammaphot{
\gamma_{\phi^\dagger\phi}^{(photon)} = - \gamma_{\phi \phi}^{(photon)} = -{ \pi^2 N_\Delta^4 g^2 q^2  \over 2\Delta-1}\,.
} 

\item{3.} Now we come to the graviton contribution.  This is
\eqn\gammagravphicphi{
\gamma_{\phi^\dagger\phi}^{(graviton)} = -{1\over2}\int d^4x\,\sqrt{-g}\,\Big( h_{\mu\nu}[\psi_0,\psi_0]T^{\mu\nu}[\psi_0^*,\psi_0^*]+h_{\mu\nu}[\psi_0,\psi_0^*]T^{\mu\nu}[\psi_0^*,\psi_0]\Big)
}
In contrast to the photon case, here $T^{\mu\nu}$ is symmetric in its arguments, so the first term no longer vanishes, and the second term is equal to the $\phi\phi$ anomalous dimension \phiphigrav. The first term corresponds to s-channel graviton exchange. To evaluate it, we need to take into account the fact that $T_{\mu\nu}[\psi_0^*,\psi_0^*]$ 
is time-dependent. So the metric response is as well:
\eqn\genfgII{\eqalign{
h_{\rho\rho} ={2 \over 3}N_\Delta^2  \Delta  (y^2 -1) y^{2\Delta -2}e^{2i\Delta t}  ,\qquad 
h_{tt} = -{2 N_\Delta^2 \Delta (\Delta-1)    \over 3 (\Delta+1)}  y^{2\Delta}e^{2i\Delta t}.
}}
Substituting this into \gammagravphicphi, we find
\eqn\VgravII{
 -{1\over2}\int d^4x\,\sqrt{-g}\, h_{\mu\nu}[\psi_0,\psi_0]T^{\mu\nu}[\psi_0^*,\psi_0^*] = -{2\pi^2N_\Delta^4 \Delta^2 (2\Delta-5)   \over 3 (\Delta-1)(2\Delta-1)(2\Delta+1)} \,.
}
Adding this to \phiphigrav, we obtain
\eqn\VgravIII{\eqalign{
\gamma_{\phi^\dagger\phi}^{(graviton)}
 &=-{2\pi^2 N_\Delta^4\Delta^2  (2\Delta^2-\Delta-7)\over 3(\Delta-1)(2\Delta-1)(2\Delta+1)}
 }}

\bigskip

Finally, let us put together all the different contributions \quarticgammaphicphi, \phicphigammaphot, and \VgravIII.
The total $\phi^\dagger\phi$ anomalous dimension is:
\eqn\phistarphigammatot{\eqalign{
\gamma_{\phi^\dagger\phi} &= \gamma_{\phi^\dagger\phi}^{(quartic)} + \gamma_{\phi^\dagger\phi}^{(photon)}
+\gamma_{\phi^\dagger\phi}^{(graviton)} 
= {2\pi^2 N_\Delta^4 \Delta\over 2\Delta-1}\left( b - {2\Delta(2\Delta+3)\over 3(2\Delta+1)}\right)
}}
In the second equation, we have again substituted \abreln\ for $a$. Thus $\gamma_{\phi^\dagger\phi}$ depends on two parameters -- $\Delta$ and $b$. A contour plot of $\gamma_{\phi^\dagger\phi}$ is shown in fig.\ 1. 
At large $\Delta$, the anomalous dimension
is always negative.  This fact can be understood physically by
noting that at large $\Delta$, the wavefunctions $\psi_0$ are very narrowly
concentrated at small $\rho$, and thus the binding energies are controlled
by the flat-space limit of AdS. In this limit, the gravitational
and electromagnetic binding energies dominate over the contact term.
Since both of these forces between two particles of opposite charge
are attractive, the binding energy is always negative at large $\Delta$.  
 
More generally, however, the anomalous dimension can take either sign.
For $b > {10 \over 9}$, it becomes positive in the range
\eqn\Deltarange{ 
 1\le \Delta < {3 \over 4} \left( \sqrt{b^2 - {2 \over 3} b +1}
 +(b-1) \right)
 }
For $b<{10\over 9}$,  $\gamma_{\phi^\dagger \phi}$ is negative for all $\Delta\ge 1$.  

\bigskip
\ifig\myfig{ Contours of $\gamma_{\phi^\dagger \phi}$ as a function of
the parameters $b$ and $\Delta$. Positive $\gamma_{\phi^\dagger \phi}$ occurs
only for $b > {10 \over 9}$, and then for $\Delta$ below a critical
value that grows with increasing $b$.  }{\epsfxsize=0.55\hsize\epsfbox{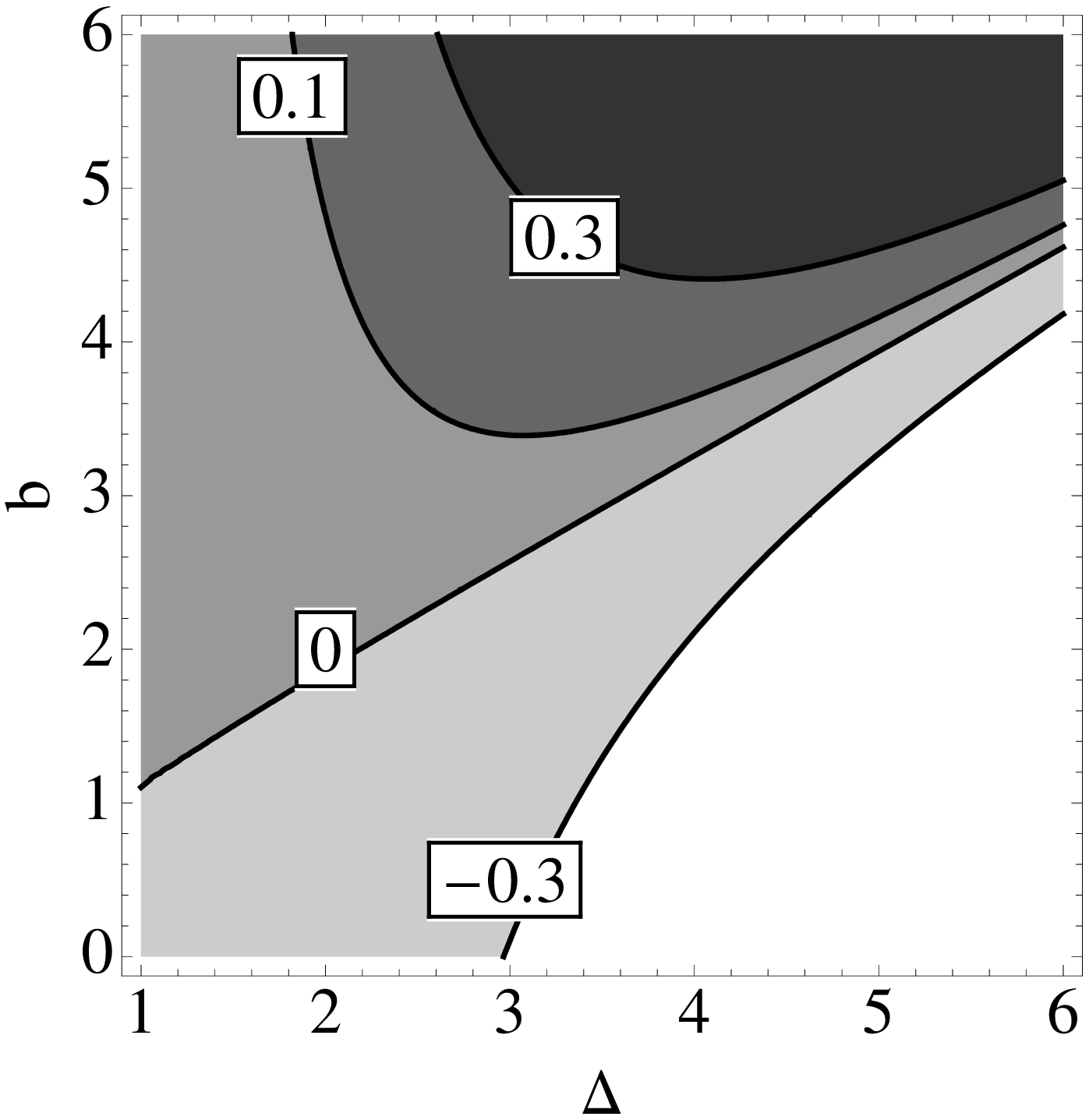}}

\newsec{Anomalous dimensions via the four point function in AdS/CFT}

\subsec{Anomalous dimensions via the four point function in general}

In this section, we will obtain the leading-order anomalous dimensions a different way, by calculating the four-point function and then taking different short distance limits where the lowest dimension operators in the $\phi\phi$ and $\phi^\dagger\phi$ OPE's dominate the expansion in conformal blocks. This will serve as a check of our new and improved method presented in the previous section. 
The more casual reader should feel free to skip over this technical section and head directly for section 5.

We start by summarizing how one calculates operator dimensions from the four point function in general CFTs. Consider the four point function $\< \CO_1(x_1) \CO_2(x_2) \CO_1^\dagger(x_3)
\CO_2^\dagger(x_4)\>$ involving two scalar primaries $\CO_1$ and $\CO_2$. We will restrict our attention
to the special case where both $\CO_1$ and $\CO_2$ have the same dimension
$\Delta$.  This four-point function is constrained by conformal symmetry to take the form:
\eqn\genfourpt{
\CC_{\CO_1\CO_2}(x_1,x_2,x_3,x_4)\equiv \< \CO_1(x_1) \CO_2(x_2) \CO_1^\dagger(x_3)
\CO_2^\dagger(x_4)\> = { \CF_{\CO_1\CO_2}(u,v) \over x_{12}^{2\Delta} x_{34}^{2\Delta} },
}
where $u$ and $v$ are the conformal cross ratios:
\eqn\uvdef{
 u\equiv {x_{12}^2x_{34}^2\over x_{13}^2x_{24}^2},\qquad v\equiv {x_{14}^2x_{23}^2\over x_{13}^2x_{24}^2}
}
It is often convenient to change variables from $u$ and $v$ to $x$ and $z$,
defined by  $u=xz$, $v=(1-x)(1-z)$.

In general, information about $\CO_1\times \CO_2$, including the dimension and OPE coefficients
of all operators appearing in the OPE, can be extracted from
 \genfourpt\ in the $x_1\to x_2$ limit.  In this limit, one has $u\to 0$ and $v\to 1$ (equivalently, $x,\,z\to 0$), and $\CF_{\CO_1\CO_2}(u,v)$ can be expanded in conformal blocks:
\eqn\Fexpcb{
\CF_{\CO_1\CO_2}(u,v) = \sum_{\CO} |C_{\CO_1\CO_2\CO}|^2 G(\Delta_\CO,\ell_\CO;x,z)
}
Here the sum is over conformal primaries $\CO$ (with dimension $\Delta_\CO$ and spin $\ell_\CO$) appearing in the OPE of $\CO_1\CO_2$, and $C_{\CO_1\CO_2\CO}$ is the OPE coefficient.\foot{The overall normalizations of the OPE coefficients are fixed by requiring all the primary operators to have canonically normalized two-point functions. For example, for scalar primaries, one has $\langle \CO^\dagger(x) \CO(0) \rangle = |x|^{-2\Delta_\CO}$.}
In \refs{\DolanUT,\DolanHV}, Dolan \& Osborn discovered a simple closed form expression for the conformal blocks.
If we take $x=\epsilon\, a$ and $z=\epsilon\, b$ with $\epsilon\to0$, we find from their general result that 
\eqn\confblockexpandeps{
 G(\Delta_\CO,\ell_\CO;x=a\epsilon,z=b\epsilon) = \epsilon^{\Delta_\CO}(a b)^{\Delta_\CO-\ell_\CO\over 2} \left({a^{{\ell_\CO}+1} - b^{{\ell_\CO}+1}\over a-b}\right) + \dots
 }
So the expansion in conformal blocks \Fexpcb\ can be thought of as an expansion in operator dimensions. 

Now let us specialize to the problem at hand: determining the leading-order anomalous dimension of the double-trace scalar operator $\CO_{min} = \CO_1\CO_2$ in an SCFT with a local, weakly-coupled gravity dual. According to \confblockexpandeps, perturbing $\Delta_\CO=\Delta_\CO^{(0)}+\kappa^2 \gamma_\CO$ gives rise to $\log u$-singular terms at leading order in $\kappa^2$: 
\eqn\FuvexpandIII{\eqalign{
\CF_{\CO_1\CO_2}(u,v)
&= {1\over2}\kappa^2 \log u\sum_{\CO}  
  \gamma_{\CO}\, |C_{\CO_1\CO_2\CO}^{(0)}|^2  \, \epsilon^{\Delta_\CO^{(0)}}(a b)^{\Delta_\CO^{(0)}-\ell_\CO\over 2} \left({a^{{\ell_\CO}+1} - b^{{\ell_\CO}+1}\over a-b}\right) +\dots
  }}
Here $C_{\CO_1\CO_2\CO}^{(0)}$ are the OPE coefficients in the free theory. But the $\CO_1\times\CO_2$ free OPE is equivalent to Taylor expansion: only double-trace operators of the schematic form
\eqn\doubletraceO{
 \CO = \CO_1 \lrpar_{\mu_1} \dots \lrpar_{\mu_\ell} (\lrpar)^{2n} \CO_2
}
appear. These have dimension and spin given by
\eqn\dimspin{
\Delta_\CO^{(0)} =  2\Delta+2n+\ell,\qquad \ell_\CO = \ell\qquad (n,\,\,\ell\ge 0)
}
in the free theory. So their dimension is bounded from below by $2\Delta$, saturated only by $\CO_{min}=\CO_1\CO_2$ with $n=\ell=0$. Therefore, in the expansion \FuvexpandIII, the leading term has $u^{\Delta}$ and corresponds to precisely the operator we are interested in. We conclude that 
\eqn\Fuvexpandii{
\CF_{\CO_1\CO_2}(u,v) 
= {1\over2}  \kappa^2   \gamma_{\CO_{\rm min}}
|C_{\CO_1\CO_2\CO_{\rm min}}|^2  \,u^{\Delta}\log u+ \dots
}
That is, by selecting out the leading $\log u$ singularity in the $\CO(\kappa^2)$ correction to the four-point function, we can infer the leading-order anomalous dimension of $\CO_1\CO_2$. 

Below, we will derive the leading-order anomalous dimensions for $(\CO_1,\CO_2)=(\phi,\phi)$ and $(\phi^\dagger,\phi)$, using \Fuvexpandii. For our purposes, we will need
\eqn\opecoeffs{
|C_{\phi, \phi, \phi \phi}^{(0)}|^2 = 2,\qquad |C_{\phi^\dagger, \phi, \phi^\dagger \phi}^{(0)}|^2 =1
}
which follows from requiring $\phi\phi$ and $\phi^\dagger\phi$ to have canonically normalized two-point functions.

\ifig\myfig{Witten Diagrams for the quartic contact interaction
(left) and photon, graviton exchange (middle, right). 
}{\epsfxsize=0.9\hsize\epsfbox{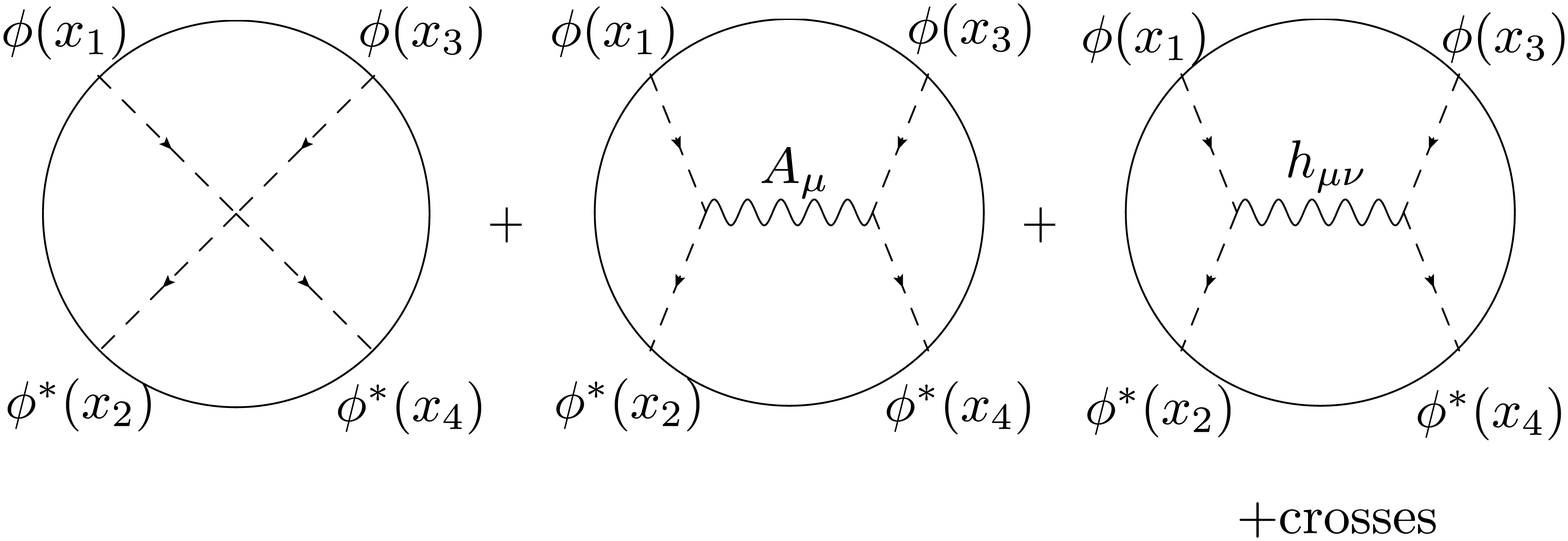}}

\subsec{AdS/CFT calculation of four-point functions}

The $\CO(\kappa^2)$ correction to the four-point function is calculated using the standard techniques of AdS/CFT.  For the model \fullaction, the relevant Witten diagrams are shown in fig.\ 2. We see that there are three types of contributions to the four-point function in general -- those from scalar quartic interactions, photon exchange, and graviton exchange. We will separate out these contributions and write
\eqn\Ccontribgen{
 \CC_{\CO_1\CO_2}(x_1,x_2,x_3,x_4) = \CC_{\CO_1\CO_2}^{(quartic)}(x_1,x_2,x_3,x_4)+\CC_{\CO_1\CO_2}^{(photon)}(x_1,x_2,x_3,x_4)+\CC_{\CO_1\CO_2}^{(graviton)}(x_1,x_2,x_3,x_4)
}

Witten diagram calculations are generally more tractable in Euclidean AdS Poincar\'e patch coordinates
\eqn\poincarecoord{
ds^2 = {d\vec w^2+(dw^0)^2\over (w^0)^2}.
} 
In what follows, we will  need the formula for the bulk-to-boundary propagator:
\eqn\bbprop{
K_\Delta(w,\vec x) =N_\Delta \left({w^0\over (w^0)^2+(\vec w-\vec x)^2}\right)^\Delta .
}
The normalization factor is fixed by imposing the canonically normalized two-point function in the CFT; it is exactly equal to the normalization of the ground-state single-particle wavefunction $\psi_0(x)$ derived in section 3.\foot{For $1 < 
\Delta < 2$, there are two available branches $\Delta_+, \Delta_-$ of the 
mass-dimension relation $m^2 = \Delta (\Delta-4)$, and one must modify
the AdS/CFT prescription for obtaining correlation functions.  There are
several equivalent procedures:
the earliest method was to use \bbprop\ and
Legendre transform the result \KlebanovTB; alternatively, one may impose
 modified boundary conditions for the bulk field in order to select
the smaller branch $\Delta_-$, in which case the bulk-to-boundary
propagator is modified \HartmanDY.  In either approach, the answers at the end of the day seem to be trivially
obtainable by ignoring all these subtleties and just analytically continuing from $\Delta>2$ to $\Delta<2$ \HartmanDY. This is what we will do here.
We note that an advantage of the
method in section 3 is that such subtleties never arise.  There, the wavefunctions
$\psi_0$ are valid for any $\Delta$ above the unitarity bound.} Note that for indices, we will be following the conventions of  \FreedmanTZ. So in general, indices will be raised, lowered, and contracted using the flat Euclidean metric $\delta_{\mu\nu}$. For instance, $w^\mu=w_\mu$. This includes the squares of coordinates, e.g. $w^2 = w^\mu w^\mu$, etc.. When the AdS metric is called for, it will be exhibited explicitly.

Now we are finally ready to derive the leading $1/N$ correction to the anomalous dimensions, using \Fuvexpandii, \opecoeffs, and \Ccontribgen, starting with the $\phi\phi$ case.

\subsec{$\phi\phi$ anomalous dimension}

As discussed in the previous subsection, the four-point function has three contributions: 
\eqn\fpfmodel{\eqalign{
& \CC_{\phi\phi}(x_1,x_2,x_3,x_4) = \CC_{\phi\phi}^{(quartic)}(x_1,x_2,x_3,x_4)+\CC_{\phi\phi}^{(photon)}(x_1,x_2,x_3,x_4)+\CC_{\phi\phi}^{(graviton)}(x_1,x_2,x_3,x_4)\cr
}}
Using the Lagrangian \fullaction, we obtain for the scalar contribution:
\eqn\scconfour{\eqalign{
& \CC_{\phi\phi}^{(quartic)} = -a \int d^5w\sqrt{g}  \,K_{\Delta}(w,\vec x_1)K_{\Delta}(w,\vec x_2)K_{\Delta}(w,\vec x_3)K_{\Delta}(w,\vec x_4)\cr
& - b \int d^5w\sqrt{g}\,g^{\mu\nu}(w)\partial_\mu K_{\Delta}(w,\vec x_1)K_{\Delta}(w,\vec x_2)\partial_\nu K_{\Delta}(w,\vec x_3)K_{\Delta}(w,\vec x_4)  + (\vec x_1\leftrightarrow \vec x_2,\vec x_3\leftrightarrow \vec x_4)  \cr
}}
It is straightforward to extract the $\log u$ divergence in \scconfour\ in the limit $x_1\to x_2$, $x_4\to \infty$. (We can always take $x_4\to \infty$ via a conformal transformation, without loss of generality.) After a shift in the integration variable $\vec w\to \vec w+\vec x_1$, the divergence comes from the region of integration where $w^0\sim |\vec w|\sim |x_{12}|\ll |x_{13}|$. Then a quick way to extract the coefficient of the log divergence is to set $x_1=x_2$ in the integrand, and cutoff the divergent integral at $|x_{12}|$. The result is
\eqn\phiphiquartic{
\lim_{x_1\to x_2}\CF_{\phi\phi}^{(quartic)} 
 =  {\pi^2\beta_\Delta^2(a+b\Delta(2-\Delta))\over (2\Delta-1)(\Delta-1)} u^\Delta \log u + \dots
}
where $\dots$ is higher order in $u$. Using \Fuvexpandii\ and \opecoeffs, we obtain the anomalous dimension quoted in \quarticgamma.\foot{One can also extract the $\log u$ divergence more properly (and painfully) using the identities for $D$ and $H$ functions listed in \DolanUT, see especially (C.8) of that paper. We have checked all our results here using this method.}

For the photon contribution:
\eqn\fpfpho{\eqalign{
& \CC_{\phi\phi}^{(photon)} = {q^2\over2!}\int d^5w\sqrt{g}\, g^{\mu\nu}(w) J_\mu(w;\vec x_2,\vec x_4) I_\nu(w;\vec x_1,\vec x_3)+ (x_1\leftrightarrow x_2,x_3\leftrightarrow x_4)
}}
Here
\eqn\JcurrentK{
J_\mu(w;x,y) = i\Big(K_{\Delta}(w,\vec x)\partial_\mu K_{\Delta}(w,\vec y) - K_{\Delta}(w,\vec y)\partial_\mu K_{\Delta}(w,\vec x)\Big)\ ,
} 
is the $U(1)$ current \JTdef\ evaluated with bulk-to-boundary propagators, and
\eqn\Imudef{\eqalign{
& I_\mu(w;\vec x_1,\vec x_3) = \int\,d^5z\sqrt{g}\,g^{\nu\rho}(z) J_\nu(z;\vec x_1,\vec x_3)G_{\rho\mu}(z,w)
}}
with $G_{\rho\mu}(z,w)$ being the massless photon propagator (explicit formulas are given in \refs{\DHokerGD,\DHokerJC}). We obtain a closed-form expression for $I_\mu(w;\vec x_1,\vec x_3)$ using the ingenious method of \DHokerNI, whereby the problem of integrating over $z$ is converted into the much easier problem of solving a differential equation in $w$. The differential equation follows from the fact that the photon propagator is a Green's function in AdS$_5$.\foot{In \DHokerNI, arbitrary spacetime dimension $d$ was considered, but attention was restricted to specific values of the operator dimension $\Delta$. Here we focus on $d=4$, and we find that $I_\mu$ can be obtained for any $\Delta$.} Solving this differential equation with appropriate boundary conditions, one finds: 
\eqn\fpfphozinv{\eqalign{
& I_\mu(w;\vec x_{1},\vec x_3) = -{iN_\Delta^2\over 2(\Delta-1)} |\vec x_{13}|^{-2\Delta} \left({(w-x_{3})^\mu\over (w-x_3)^2}-{(w-x_1)^\mu \over (w-x_{1})^2}\right) f(t) \cr
}}
where
\eqn\ftdef{\eqalign{
& f(t) \equiv {t(1-t^{\Delta-1})\over 1-t}\cr
& t \equiv { (w^0)^2\vec x_{13}^2\over (w-x_1)^2 (w-x_3)^2}
}}
This is explained in more detail in appendix B. 

To extract the $\log u$ divergence from \fpfpho\fpfphozinv\ftdef, we must first expand $f(t)$ in powers of $t$:
\eqn\foftseries{
f(t) = \sum_{k\ge 1}t^k-\sum_{k\ge 1}t^{k+\Delta-1}
}
Using the fact that $t^\alpha \sim |\vec x_{13}|^{2\alpha} K_\alpha(w,\vec x_1)K_\alpha(w,\vec x_3)$ (up to normalization), the $\log u$ divergence from each term in the series representation of $f(t)$ can be extracted as in the scalar case above. The answer for a given power of $t$ is:
\eqn\phiphiphota{
\lim_{x_1\to x_2}\CF_{\phi\phi}^{(photon)} \Big|_{t^\alpha}
 = -{2\pi^2N_\Delta^4 q^2\Delta\over (\alpha+\Delta-1)(\alpha+\Delta)} u^\Delta \log u+\dots
}
Substituting this in \foftseries, we obtain the photon anomalous dimension quoted in \phiphigammaphot.

Finally, the graviton contribution is:
\eqn\fpfgrav{
 \CC_{\phi\phi}^{(graviton)} = {1\over2!\cdot 4} \int d^5w\sqrt{g}\, g^{\mu\rho}(w)g^{\nu\lambda}(w)T_{\mu\nu}(w;\vec x_2,\vec x_4) I_{\rho\lambda}(w;\vec x_1,\vec x_3)+ (x_1\leftrightarrow x_2,x_3\leftrightarrow x_4)
}
Here
\eqn\TgravK{\eqalign{
T_{\mu\nu}(w;x,y) &= \Big( \partial_\mu K_{\Delta}(w,\vec x)\partial_\nu K_{\Delta}(w,\vec y)+(\mu\leftrightarrow \nu)\Big)\cr
&\qquad -g_{\mu\nu}\Big(g^{\rho\lambda}(w)\partial_\rho K_{\Delta}(w,\vec x) \partial_\lambda K_{\Delta}(w,\vec y)+m^2K_{\Delta}(w,\vec x)K_{\Delta}(w,\vec y)\Big)
}}
is the stress-energy tensor \JTdef\ evaluated on the bulk-to-boundary propagators, and
\eqn\Imunudef{
 I_{\mu\nu}(w;\vec x_1,\vec x_3)= \int d^5z\sqrt{g}\,g^{\mu'\rho'}(z)g^{\nu'\lambda'}(z) T_{\mu'\nu'}(z;\vec x_1,\vec x_3)G_{\rho'\lambda'\mu\nu}(z,w)
}
with $G_{\rho'\lambda'\mu\nu}(z,w)$ being the graviton propagator (explicit formulas in \refs{\DHokerJC}). 
Using again the methods of \DHokerNI, we obtain for $I_{\mu\nu}$:
\eqn\fpfgravz{\eqalign{
I_{\mu\nu}(w;\vec x_{1},\vec x_3) & = {N_\Delta^2 \Delta\over (\Delta-1)} |\vec x_{31}|^{-2\Delta}  {1\over (w^0)^2}\left( 
{1\over3}\delta_{\mu\nu} -J_{\mu0}(w-x_1)J_{\nu0}(w-x_1)\right) f(t)\cr
 & \qquad \qquad +\,\,( {\rm gauge-dependent})
}}
where $J_{\mu\nu}(w) = \delta_{\mu\nu}-2w^\mu w^\nu/w^2$ is the inversion tensor. Note that \fpfgravz\ is not symmetric in $x_1\leftrightarrow x_3$. This is because we have neglected gauge-dependent terms in evaluating the $z$ integral. These must drop out after doing the $w$ integral.

Repeating the same manipulations as for the scalar and photon cases, we obtain
\eqn\phiphigrava{
\lim_{x_1\to x_2}\CF_{\phi\phi}^{(graviton)} \Big|_{t^\alpha}
 = {4\pi^2N_\Delta^4 \Delta^2(\Delta(\Delta-4)+\alpha(\Delta+2))\over 3(\alpha+\Delta)(\alpha+\Delta-1)(\alpha+\Delta-2)} u^\Delta \log u+\dots
}
Substituting this in \foftseries, we obtain the graviton anomalous dimension quoted in \VgravI.

\subsec{$\phi^\dagger\phi$ anomalous dimension}

Next we consider the $\phi^\dagger\phi$ anomalous dimensions. These are more complicated, because generally they have contributions from $s$-channel diagrams as well as $t$, $u$ channel diagrams. Here it makes sense to separate out the $s$-channel contribution, because the $t$ and $u$ channel pieces have already been computed above. 

The four-point function relevant to $\phi^\dagger\phi$ is easily obtained from that for $\phi\phi$:
\eqn\fpfrel{
\CC_{\phi^\dagger\phi}(x_1,x_2,x_3,x_4) = \CC_{\phi\phi}(x_1,x_3,x_2,x_4) 
}
Applying this to \scconfour, and extracting the $\log u$ divergence as above, we find
\eqn\phicphiquartic{
\lim_{x_1\to x_2}\CF_{\phi^\dagger\phi}^{(quartic)}
 = \left( {(a+b\Delta^2)\over 2(2\Delta-1)(\Delta-1)}  + {(a+b\Delta(2-\Delta))\over 2(2\Delta-1)(\Delta-1)} \right) \pi^2N_\Delta^4 u^\Delta \log u + \dots
}
The first term is the $s$-channel contribution; the second term is the contribution of the $t+u$ channels, which we have  computed above.

For the photon interaction, we obtain the opposite of the $\phi\phi$ case \phiphiphota, as expected (the factor of 2 difference is due to \opecoeffs) :
\eqn\phicphiphoton{
\lim_{x_1\to x_2}F_{\phi^\dagger\phi}^{(photon)} \Big|_{t^\alpha}
 = {q^2\Delta\over (\alpha+\Delta-1)(\alpha+\Delta)} \pi^2N_\Delta^4  u^\Delta \log u+\dots
}
Here the $s$-channel contribution is found to vanish -- essentially there is an extra suppression as $x_1\to x_2$ from $J_\mu(w;\vec x_1,\vec x_2)\to 0$ in \fpfpho. Meanwhile, the $t+u$ channels differ by a sign from \phiphiphota, which is a consequence of $J_\mu(w;x,y)=-J_\mu(w;y,x)$. This all agrees perfectly with \phicphigammaphot.

Finally, for the graviton interaction, we find
\eqn\phicphigraviton{\eqalign{
\lim_{x_1\to x_2}\CF_{\phi^\dagger\phi}^{(graviton)} \Big|_{t^\alpha}
& =  \left(-\delta_{\alpha\Delta} { \Delta^2(2\Delta-5)\over 3(2\Delta-1)(2\Delta+1)}+
 { \Delta^2(\Delta(\Delta-4)+\alpha(\Delta+2))\over 3(\alpha+\Delta)(\alpha+\Delta-1)(\alpha+\Delta-2)}\right)\cr
 &\qquad\times  \pi^2N_\Delta^4 u^\Delta \log u+\dots
 }}
Here the $s$-channel only contributes for a certain power of $t$; the other channels are again the same as above. The anomalous dimension obtained from \phicphigraviton\ is in perfect agreement with \VgravIII.

\newsec{Example: $SU(2,1)/U(2)\times U(1)$ coset}

Here we will study an explicit example of $\CN=2$, $d=5$ gauged supergravity with one hypermultiplet and no vector multiplets. This is the sigma model whose target space is the quaternionic K\"ahler manifold
\eqn\unitarywolf{
\CM={SU(2,1)/( SU(2)\times U(1))}
} 
In $d=4$, this sigma model corresponds to the ``universal hypermultiplet" (containing the axion-dilaton) of type II compactification on a general Calabi-Yau threefold \CecottiQN. In $d=5$, it similarly describes the axion-dilaton sector of general type II compactifications on $AdS_5\times {\rm SE}_5$ \refs{\CassaniUW\LiuSA\GauntlettVU-\SkenderisVZ}. This space is studied in fairly explicit detail in the literature; we have found \refs{\CeresoleWI,\deWitBK,\GennipThesis}\ to be especially useful. Generalizations to $n$ hypermultiplets exist, and it would be interesting to study these as well.

A convenient representation of the coset is:
\eqn\Lcosetexample{
L(z) = {1 \over 1-z^\dagger z} \pmatrix{ {\bf 1}_2 (1-z^\dagger z) + 2 z z^\dagger  & 2 z \cr 2 z^\dagger & 1+z^\dagger z } \in SU(2,1)
}
where $z=(z^1,z^2)$ are complex coordinates on $\CM$.\foot{The complex coordinates used here are related to the real coordinates $(V,\sigma,\theta,\tau)$ in \CeresoleWI\ via 
\eqn\Vtoz{
V = {1-z^\dagger z\over (1+z^1)(1+\bar z^{\bar 1})},\quad \sigma =-2{\rm Im}({z^1\over 1+z^1}),\quad \theta = {\rm Re}({z^2\over 1+z^1}),\quad \tau = -{\rm Im}({z^2\over 1+z^1})
}}  
The unique left- and right-invariant metric on $\CM$ is inherited from 
\eqn\naturalmetric{
ds^2={1\over4}\Tr\,L^{-1}dL L^{-1} dL = -{1\over4}\Tr\,\eta dL^\dagger \eta dL
}
up to multiplication by a constant. (Here the constant is fixed to agree with the conventions in the literature, e.g.\ \CeresoleWI.) 
Explicitly we have:
\eqn\metricz{
ds^2=2g_{i\bar j}dz^i d\bar z^{\bar j},\qquad g_{i\bar j} = {\bar z^{\bar i} z^{j}\over (1-z^\dagger z)^2} + { \delta_{i j}\over (1-z^\dagger z)},
}
which comes from the following K\"ahler potential:
\eqn\kpotz{
\CK = -\log (1-z^\dagger z )
}

The manifold $\CM$ has an $SU(2,1)$ isometry group. 
The  Killing vectors $\delta_a z^i = K_a\,^{i}$ ($a=1,\dots,8$) are:
\eqn\kvsol{\eqalign{
& K_1 = -i\pmatrix{z^2\cr z^1},\quad K_2=\pmatrix{-z^2\cr z^1},\quad K_3=i\pmatrix{-z^1\cr z^2},\quad K_4=i\pmatrix{z^1\cr z^2}\cr
& K_5=\pmatrix{1-(z^1)^2\cr -z^1z^2},\quad K_6= i\pmatrix{1+(z^1)^2\cr z^1 z^2},\quad K_7=\pmatrix{-z^1z^2\cr 1-(z^2)^2},\quad K_8=i\pmatrix{z^1z^2\cr 1+(z^2)^2}
}}
Note that $K_{1,2,3}$ generates an $SU(2)$ subgroup of $SU(2,1)$, and  $K_4$ generates a trivial $U(1)$ subgroup.

As discussed above, the hypermultiplet interactions arise from gauging the isometries \kvsol. That is, we choose an isometry generated by some linear combination of \kvsol,
\eqn\kvchoose{
K = c_a K_a
}
This determines the hypermultiplet Lagrangian
\eqn\scalarLquarticgen{\eqalign{
 & \CL = -g_{i\bar j} D_\mu z^i D^\mu \bar z^j - V(z,\bar z) \cr
}}
via
\eqn\covderhyperK{
D_\mu z^i = \partial_\mu z^i + A_\mu K^i
}
and
\eqn\Vhyper{
V(z,\bar z) = {3\over2}K^i \bar K_{i} -{1\over4}(D^i \bar K^{\bar j}-D^{\bar j} K^{i})(g_{i\bar j}g_{k\bar \ell}-2g_{i\bar \ell}g_{k\bar j})(D^k \bar K^{\bar \ell}-D^{\bar \ell} K^{k})
}
Here indices are being raised and lowered with the K\"ahler metric \metricz. This formula for the scalar potential is derived in appendix C.

According to the hyperino SUSY transformation, the SUSY vacua are located at the solutions to $K^i=0$ \AndrianopoliCM.
Since the target space is a symmetric space, we lose no generality by assuming that the vacuum is located at $z^1=z^2=0$. Then the isometries that can be gauged are $K_{1,2,3,4}$ from \kvsol.
Via an $SU(2)$ rotation, we also lose no generality if we only take $K_3$ out of $K_{1,2,3}$ to be gauged. Thus the most general isometry that can be gauged is: 
\eqn\Ktake{
K=\sqrt{3\over 2} (K^3 +c K^4 )
}
for an arbitrary real constant $c$, whose physical significance will
become apparent shortly. The overall normalization in \Ktake\ is chosen so that the AdS radius $R=1$. Substituting \Ktake\ into \scalarLquarticgen, \covderhyperK, and \Vhyper, and expanding around $z^i=0$, we obtain the hypermultiplet Lagrangian to quartic order:
\eqn\scalarLquartic{\eqalign{
\CL  & =  -D_\mu z^1D^\mu \bar z^1  - D_\mu z^2D^\mu \bar z^2 +6 - m_+^2 |z^1|^2 - m_-^2|z^2|^2 \cr
 & - \Bigg((2|z^1|^2+|z^2|^2)\partial_\mu z^1\partial^\mu\bar z^1 +(|z^1|^2+2|z^2|^2)\partial_\mu z^2\partial^\mu \bar z^2+z^2\bar z^1\partial_\mu z^1\partial^\mu \bar z^2  +z^1\bar z^2\partial_\mu z^2  \partial^\mu \bar z^1 \Bigg)\cr
 & +3\Bigg( (1-c^2)|z^1|^4 + (3- 2c^2)|z^1|^2|z^2|^2 + (1-c^2)|z^2|^4\Bigg) + \dots
 }}
Here: 
\eqn\gaugecouplingK{
\pmatrix{ D_\mu z^1 \cr D_\mu z^2} =  \pmatrix{ \partial_\mu z^1+i\sqrt{3\over2}A_\mu(c-1)z^1 \cr \partial_\mu z^2+i\sqrt{3\over2}A_\mu(c+1 )z^2}
}
the masses are given by:
\eqn\mpm{
m_\pm^2=\left({3c\over2}\mp {3\over2}\right)\left({3c\over2}\pm {5\over2}\right)
}
the second line of \scalarLquartic\ comes from expanding out the target space metric, and  the third line comes from expanding out the scalar potential.
From the AdS/CFT dictionary \gfixed, \mDeltarel, and \susyq, we conclude that $z^2$ is dual to a chiral primary $\phi$ with dimension and R-charge given by
\eqn\Deltaofphi{
 \Delta_\phi={3c\over2}+{3\over2},\qquad  R_\phi={2\over3}\Delta_\phi
 }
while $z^1$ is dual to its $F$-component with dimension and R-charge given by
\eqn\DeltaofFphi{
\Delta_{F_\phi}=\Delta_\phi+1,\qquad  R_{F_\phi}={2\over3}\Delta_\phi-2
} 
Other branches of the mass-dimension relation do not obey $R={2\over3}\Delta$ for the chiral primary, and therefore are not consistent with supersymmetry.

 The quartic interactions for the chiral primary $z^2$ are:
\eqn\Lintcp{
\CL_{quartic} = -2|z^2|^2\partial_\mu z^2\partial^\mu \bar z^2 +3(1-c^2)|z^2|^4
}
In terms of our earlier parameterization \scalarpot, this sets the
potential coefficients to be
\eqn\cosetpot{
b  = 2, \ \ \ \ \ a = -3(1-c^2) = {4 \over 3}\Delta (\Delta-3) .
}
This does indeed satisfy the relation \abreln, as required by supersymmetry. Correspondingly, the $\phi\phi$ anomalous dimension vanishes to this order. Meanwhile, the $\phi^\dagger\phi$ anomalous dimension is easily read off from \phistarphigammatot:
\eqn\cosetgamma{
\gamma_{\phi^\dagger\phi} 
 = - { 4\pi^2N_\Delta^4\Delta \left( 2 \Delta^2 - 3 \Delta -3 \right) 
\over 3 (2\Delta-1)(2\Delta+1) }.
}
This is negative for large $\Delta$, but crosses over to positive for $\Delta\lesssim 2.2$.

\newsec{Summary and Future Directions}

In this paper, we developed a new method for computing
anomalous dimensions of double-trace operators in 4d $\CN=1$ SCFTs with local, weakly-coupled AdS$_5$ supergravity duals, at leading order in the $1/N$ approximation. Anomalous dimensions are dual to binding energies of two-particle states in the bulk. By directly computing these binding energies in the Hamiltonian formulation of AdS/CFT, we have considerably simplified previous indirect approaches based on the four-point function. 

For the sake of concreteness and simplicity, we have focused here on a minimal effective model consisting of a single complex scalar $\phi$ coupled supersymmetrically to gravity and the graviphoton in AdS$_5$. We  calculated the anomalous dimensions of $\phi\phi$ and $\phi^\dagger\phi$, using both our new method and the four-point function method. We  found complete agreement between the two methods, which provides a strong check of our results. Using the fact that $\gamma_{\phi\phi}$ must be zero in any $\CN=1$ SCFT, we  derived a new constraint \abreln\ on the parameters of the effective theory. Although we have focused on the AdS duals of supersymmetric CFTs in this paper, our general techniques and results clearly apply equally well, with minor modifications, to the AdS duals of non-supersymmetric CFTs. 

Our result \phistarphigammatot\ for the anomalous dimension of $\phi^\dagger\phi$ in our minimal model is illuminating. We found that depending on the parameters of the model, the anomalous dimension can be positive or negative. That is, we can have either $\Delta_{\phi^\dagger \phi} > 2 \Delta_\phi$ or $\Delta_{\phi^\dagger\phi} < 2\Delta_\phi$. In section 5, we  considered a specific supergravity model (the ``universal hypermultiplet") which is a special case of our minimal toy model, and which confirms that both signs of the anomalous dimensions are possible.

This has interesting consequences for what can be proved about $\gamma_{\phi^\dagger\phi}$ using general CFT principles such as crossing symmetry and unitarity. In the  framework of effective AdS/CFT, correlation functions obtained at any fixed order in the $1/N$ expansion  automatically satisfy crossing symmetry and unitarity. Thus even without a UV completion, any results derived in an effective AdS/CFT setup are guaranteed to be compatible with any and all bounds extracted using crossing symmetry and unitarity alone. The results in this paper,
for example, robustly demonstrate that $\gamma_{\phi^\dagger \phi} >0$ is
consistent with crossing symmetry and unitarity (and supersymmetry) of the four-point function. 

The present paper is just the first step in a much broader research programme.
Clearly, a more exhaustive study going beyond the minimal model \fullaction\ is needed. For instance, one could consider models with more hypermultiplets, and also models with massive supergravity modes, which can arise in realistic string compactifications (see e.g.\ \refs{\CeresoleZS,\CeresoleHT} and \refs{\CassaniUW\LiuSA\GauntlettVU-\SkenderisVZ}). It would be also interesting to take our methods beyond leading order, to understand the effects of loops in the effective theory, as well as $\alpha'$-suppressed higher derivative corrections. Including such modes and effects will lead to many more contributions to binding energies, and  it will be fascinating to
compare the full range of such generalizations with the bounds on
various SCFT quantities in the literature.  Bounds on quantities
other than anomalous dimensions should be explored as well; although
we have focused here on anomalous dimensions, SCFTs also face
bounds on central charges and OPE coefficients.

While the positive anomalous dimensions we have obtained here are certainly tantalizing and suggestive, there is no guarantee that our setup \fullaction\ (with general $b$ and $\Delta$) can always be UV-completed.
Thus we cannot claim to have constructed the first existence proof of positive anomalous dimensions in SCFT. At best we have an almost-existence proof.  Clearly, UV completions in string theory are sorely needed.
A promising direction here would be to study consistent truncations of type IIB string theory compactified on Sasaki-Einstein manifolds. The papers of \refs{\CassaniUW\LiuSA\GauntlettVU-\SkenderisVZ} should prove useful for this purpose. 

Alternatively, it is also conceivable that while  positive anomalous dimensions are possible in effective theories such as the one studied here, such effective theories are never realized in string theory. Along these lines, it would be interesting to see if one could deduce nontrivial constraints on the parameters of the effective theory using general principles of QFT, as was done in  \AdamsSV. 
Such constraints usually derive from analyzing the theory in a background
of non-zero $\phi$ field configurations. Most of the region of parameter space with positive anomalous dimension has $b$ positive and large,  which 
naively seems to be good for stability, 
since it behaves like a correct-sign kinetic term at $\langle \phi \rangle 
\ne 0$.  

Ultimately, one would also like to take advantage of these AdS constructions
and apply them to supersymmetric model-building. As discussed in the introduction, SCFTs with positive anomalous dimensions have several interesting applications to the hidden sector of SUSY-breaking theories, where they are used to
 suppress undesired K\"ahler potential terms under RG evolution.  For instance,
scalar mass terms in the MSSM generated at the Planck scale generically lead to large  
flavor violation at low scales.  Positive anomalous dimensions for the SUSY-breaking field can solve this problem by
causing such scalar masses to flow to zero, relative to gaugino
mass terms, restoring the Standard Model flavor-breaking structure. 
This is essentially the proposal of gaugino mediation \refs{\KaplanAC\ChackoMI-\SchmaltzGY}. (In fact, the Planck-scale is inessential for the {\it spectrum} of gaugino mediation; positive anomalous dimensions can lead to the same type of spectrum also in low-scale SUSY breaking models \DumitrescuHA.)  One can similarly imagine using positive anomalous dimensions to suppress $B_\mu$ relative to $\mu$ in gauge mediation, thereby solving the $\mu/B_\mu$ problem \refs{\DineDV\RoyNZ\CohenQC-\PerezNG}.
In all such applications, an actual calculable example of an SCFT  with the desired properties is currently lacking. Perhaps the AdS/CFT approach employed here could one day lead to such an example. Of course, a complete model along these lines would require incorporating dynamical SUSY breaking into the hidden sector in the
AdS description, and coupling it to the MSSM. How to do this in general is interesting to contemplate.

\bigskip

\noindent {\bf Acknowledgments:}

We would like to thank A.~Dymarsky, D.~Green, T.~Hartman, J.~Kaplan, J.~Liu, G.~Moore, J.~Penedones, A.~Royston, M.~Schmaltz, P.~Szepietowski, Y.~Tachikawa,  and especially J.~Maldacena for useful discussions. The research of ALF is supported in part by DOE grant DE-FG02-01ER-40676 and NSF CAREER grant PHY-0645456. The research of DS is supported
in part by a DOE Early Career Award.

\appendix{A}{Details of the binding energy calculations}

In this appendix, we will provide the detailed solution of the equations of motion \photgraveom\ in the $\phi\phi$ and $\phi^\dagger\phi$ binding energy calculations of section 3. 

First, we begin with the photon equation of motion \photgraveom\ in the $\phi\phi$ case. The starting point is the source \chargedensityII, which we repeat here for convenience:
\eqn\chargedensityIIagain{\eqalign{
J^0[\psi_0(x),\psi_0^*(x)] = - 2  \Delta  q N^2_\Delta \cos^{2\Delta +2} \rho , \ \ \ \ \  J^i[\psi_0(x),\psi_0^*(x)] =0.
}}
Due to the high degree of symmetry of the wavefunctions, we may take as an
ansatz that $A_0$ depends
only on the radial coordinate $\rho$, so the equation of motion $D_\mu F^{\mu 0 } = J^0$ simplifies dramatically: 
\eqn\potleomiiII{
\left( {y^5  \over (1-y^2)} \right) \partial_y
\left( { (1-y^2)^2  \over y} \partial_y A_0 \right) = 
 2 \Delta  q N^2\Delta  y^{2 \Delta +2 } ,
}
where $y=\cos \rho$. This can be trivially solved for $A_0$, yielding the result \potlfinsoln\ quoted in the body of the paper. Here the two integration constants are fixed by the conditions
that the potential die off at $\rho \rightarrow {\pi\over2}$ and that
it be smooth at the origin $\rho \rightarrow 0$. 

Next we consider the graviton equation of motion in the $\phi\phi$ case. Here the starting point is the stress tensor source \enmomII, which we again repeat for convenience:
\eqn\enmomIIagain{
T^\mu_{\ \nu}[\psi_0(x),\psi_0^*(x)] = 2\Delta N^2_\Delta  \cos^{2\Delta} \rho \cdot  {\rm diag} \left( 2-\Delta,
2, 2 - \Delta \sin^2 \rho, \dots, 2-\Delta \sin^2 \rho \right).
}
The symmetry of the source again suggests a very symmetric ansatz for the metric:
\eqn\metansatz{\eqalign{
g_{\mu\nu} & = g^{\rm (AdS)}_{\mu\nu} + h_{\mu\nu}(y) \cr
}}
with  only $h_{tt}$ and $h_{\rho\rho}$ nonzero. Linearizing in $h_{\mu\nu}$, \photgraveom\ has only two independent equations of motion: $G^t_{\ t} = T^t_{\ t},
G^\rho_{\ \rho} = T^\rho_{\ \rho}$ (the remaining components of Einstein's 
equation are linear combinations of these and their derivatives).  
These two are respectively,
\eqn\GlinII{\eqalign{
& {3 \over 2} y h_{\rho\rho}'(y)+{3 \over y^2-1} h_{\rho\rho}(y)  =-2 \Delta(\Delta-2) N^2_\Delta  y^{2\Delta -2}, \cr
& {3 \over 2} y h_{tt}'(y) - {3 (y^2 -2) \over y^2 -1} h_{\rho\rho}(y) +3 h_{tt}(y)  =
   4 \Delta N^2_\Delta  y^{2\Delta -2} .
}}
These are again easily solved, and yield the formulas \fgfinalII\ quoted in the text. Again, the integration constants have been fixed by demanding that the solution be regular at $y=1$ and die off at $y=0$. 

Finally, we come to the graviton equation of motion in the $\phi^\dagger\phi$ case, with the time-dependent source $T_{\mu\nu}[\psi_0,\psi_0]$. Now we modify our ansatz for the metric:
\eqn\schanans{\eqalign{
g_{\mu\nu} & = g^{\rm (AdS)}_{\mu\nu} + e^{2i\Delta t}h_{\mu\nu}(y), \cr
}}
with again only $h_{tt}$ and $h_{\rho\rho}$ nonzero. The components of Einstein's equation corresponding to $G^t_t$ and 
$G^\rho_\rho$ are now
\eqn\schaneomII{\eqalign{
3 y^3 h_{\rho\rho}'(y)(y^2-1)+ 6 y^2 h_{\rho\rho}(y) - N^2_\Delta  (y^2 -1)(4y^{2\Delta} \Delta (2+\Delta(y^2-1))) & = 0,
\cr
3 y^3 h_{tt}'(y) + 6 y^2 (h_{tt}(y) -  h_{\rho\rho}(y) {y^2 -2\over y^2 -1} ) + 4 N^2_\Delta  y^{2\Delta}\Delta
(\Delta y^2 -2) & = 0 .
}}
These are solved by \genfgII.

\appendix{B}{Details of the four-point function calculations}

Here we will review the clever technique of \DHokerNI\ for evaluating the $z$ integrals in the calculation of AdS four-point functions \Imudef\ and \Imunudef. Keep in mind our conventions for the indices, discussed below \bbprop.

After a shift to $y = w-x_1$, \Imudef\ becomes
\eqn\Imudefagain{\eqalign{
& I_\mu(y;0,\vec x_{31}) = \int\,d^5z\sqrt{g}\,g^{\nu\rho}(z) J_\nu(z;0,\vec x_{31})G_{\rho\mu}(z,y) 
}}
Next, we perform an inversion on the RHS of $I^\mu(y;0,\vec x_{31})$ above. This yields
\eqn\Iinversion{
 I_\mu(y;0,\vec x_{31}) =|x_{31}|^{-2\Delta}{1\over y^2}J_{\mu\nu}(y)I_\nu(y'-x_{31}')
}
where $J_{\mu\nu}(y) = \delta_{\mu\nu}-2y^\mu y^\nu/y^2$ is the inversion tensor, $y'^\mu={y^\mu\over y^2}$, and 
\eqn\phointstart{
I_\mu(x) = i N_\Delta^2 \int {d^{5}z\over (z^0)^{5}}g^{\nu\rho}(z) \left((z^0)^{\Delta}\partial_{\nu}\left({z^0\over  z^2}\right)^{\Delta}-\left({z^0\over z^2}\right)^{\Delta}\partial_{\nu}(z^0)^{\Delta}\right)G_{\rho\mu}(z,x)
}
The point of these manipulations is that we have reduced the $z$ integral in \Imudefagain\ to  a function of a single variable $I_\mu(x)$. Lorentz invariance, dimensional analysis, and current conservation imply:
\eqn\phointii{
I_\mu(x) =  i N_\Delta^2 {x^\mu\over x^2} g(t),\quad t\equiv {(x^0)^2\over x^2}
}
One can check that substituting $x=y'-x_{31}'$, one obtains $t$ as in \ftdef. Finally, $g(t)$ satisfies the following differential equation, derived from the fact that the photon propagator is a AdS Green's function:
\eqn\fdiffeq{
2 t^2 (t-1)g'' + 4 t^2g' = - \Delta t^\Delta.
}
After imposing appropriate boundary conditions (namely that $g(t)$ is zero at $t=0$ and smooth at $t=1$), we obtain the solution $g(t)=-{1\over 2(\Delta-1)}f(t)$ with $f(t)$ given in \ftdef. Substituting this into \phointii, and then into \Iinversion, we obtain the result \fpfphozinv\ quoted in the text.

Next, consider the same shift and inversion applied to the graviton $z$ integral \Imunudef. This yields:
\eqn\fpfgravzappendix{\eqalign{
I_{\mu\nu}(y;0,\vec x_{31})&= \int d^5z\sqrt{g}\, g^{\mu'\rho'}(z)g^{\nu'\lambda'}(z) T_{\mu'\nu'}(z;0,\vec x_{31})G_{\rho'\lambda'\mu\nu}(z,y)\cr
 &= |x_{31}|^{-2\Delta} {1\over y^4} J_{\mu\rho}(y)J_{\nu\lambda}(y)I_{\rho\lambda}(y'-x_{31}')
}}
where
\eqn\gravintstart{\eqalign{
I_{\mu\nu}(x) &= N_\Delta^2 \int {d^{5}z\over (z^0)^{5}}\,g^{\mu'\rho'}(z)g^{\nu'\lambda'}(z)\Bigg[ 2\partial_{\mu'} (z^0)^{\Delta} 
\partial_{\nu'}\left( {z^0\over z^2}\right)^{\Delta} \cr
&\qquad\qquad - g_{\mu'\nu'}\left(\partial_{\kappa'} (z^0)^\Delta \partial^{\kappa'}\left( {z^0\over z^2}\right)^{\Delta} + m^2 (z^0)^{\Delta}\left( {z^0\over z^2}\right)^{\Delta}\right)\Bigg]G_{\rho'\lambda'\mu\nu}(z,x)
}}
According to D'Hoker et al, $I_{\mu\nu}(x)$ takes the form
\eqn\Imunuform{
I_{\mu\nu}(x) = 2N_\Delta^2\Bigg[ g_{\mu\nu} h(t) + {\delta_{0\mu}\delta_{0\nu} \over (x^0)^2}\phi(t) + D_\mu D_\nu X(t) +D_{ \{ \mu}\left( {\delta_{\nu\} 0} \over x^0} Y(t)\right)\Bigg]
}
Here $X$ and $Y$ are gauge artifacts that drop out of the final integral over $w$. Differential equations analogous to \fdiffeq\ can be derived for $h$ and $\phi$, leading to:
\eqn\hphisolfinal{
h(t)=-{1\over3}\phi(t) = -{\Delta\over 3}g(t)
}
Using \hphisolfinal\ in \fpfgravzappendix\ and \Imunuform, we arrive at the result \fpfgravz\ quoted in the body of the paper.

\appendix{C}{Calculation of the Hypermultiplet Potential}

In this appendix, we will derive the formula \Vhyper\ for the hypermultiplet scalar potential. Our starting point is the general formula for the hypermultiplet potential valid for any quaternionic K\"ahler (QK) manifold. To write down this formula, we must first review some basic facts about QK manifolds. 

Let $\CM_{4n_H}$ denote a QK manifold of real dimension $4n_H$, with real coordinates $q^x$, $x=1,\dots,4n_H$. Since $\CM_{4n_H}$ has  holonomy contained within $Sp(n_H)\times SU(2)$, it is convenient to pass from the curved coordinates $q^x$ to flat coordinates $\in Sp(n_H)\times SU(2)$ via the vielbein $f^{x}_{A\alpha}$,  with $A=1,\dots,2n_H$ and $\alpha=1,2$. Let $\omega_\alpha^\beta = \omega_{x\alpha}^\beta dq^x$ denote the $SU(2)$ spin connection.  Explicitly, we have\foot{Indices $x$, $y$, \dots\ are raised and lowered  with the target space metric $g_{xy}$; indices $\alpha$, $\beta$, \dots\ are raised and lowered with $\epsilon_{\alpha\beta}$; and indices $A$, $B$, \dots\ are raised and lowered with the invariant two-tensor $C_{AB}$  on $Sp(n_H)$.}
\eqn\sutwospinconn{
  \omega_{x\alpha}^\beta =  -{1\over2}(\Gamma^{y}_{xz} f^{\beta}_{y A}f^{zA}_\alpha + f^y_{A \alpha}\partial_x f^{A \beta}_y)
  }
We can change to an explicit adjoint-valued basis with $\omega^{r=1,2,3}=-{i\over2}\omega_{\alpha}^{\beta}(\sigma^r)_{\beta}\,^{\alpha}$. 
The $SU(2)$ curvature is given by  
\eqn\sutwocurvature{
\CR^{r}=d\omega^r-\epsilon^{rst}\omega^s\wedge \omega^t 
}
From this, we construct the {\it quaternionic prepotentials} $P^{r=1,2,3}$. They are given by the formula
\eqn\prepot{
P^r = {1\over 2} D^x K^{y}\CR^{r}_{xy}
}
Finally, we have all the ingredients we need to write down the general formula for the hypermultiplet potential. It is:
\eqn\VhyperQK{
V = {3\over4} K^x K_x - 4P^r P^r = {3\over4} K^x K_x - D^{x}K^y D^{w}K^z \CR^{r}_{xy}\CR^r_{wz} 
}

Now, we would like to specialize this formula to the case of interest, $\CM_{4(n_H=1)} = SU(2,1)/SU(2)\times U(1)$, which is also a K\"ahler manifold. Passing from the real coordinates $q^x$ to the complex coordinates $z^i$, $\bar z^{\bar i}$, the first term in \VhyperQK\ can be written ${3\over2}K^i \bar K_i$. To simplify the second term in \VhyperQK, we first notice that in this case, the only nonzero components of $\CR^{r}_{xy}$ are $\CR^{r}_{i\bar j}=-\CR^{r}_{\bar j i}$. Moreover, we find that
\eqn\CRident{
\CR^{r}_{i \bar j}\CR^{r}_{k\bar \ell} = {1\over4}(g_{i\bar j}g_{k\bar \ell} - 2g_{i\bar\ell}g_{k\bar j})
}
Substituting this into \VhyperQK, we obtain \Vhyper.

\listrefs
\end